\documentclass[usegraphicx,usedcolumn,usenatbib]{mn2e}
\usepackage{txfonts}
\usepackage{color,epsfig,amssymb,longtable}
\usepackage{array,colortbl,lscape}
\usepackage{color}

\definecolor{dgreen}{rgb}{0,.5,.1} 
\definecolor{pink}{rgb}{.9,.2,.5}  
\definecolor{orange}{rgb}{.9,.4,0} 
\definecolor{darkred}{rgb}{.545,0.0,.0}

\newcommand{\halpha}{H$_{\alpha}$}
\newcommand{\hbeta}{H$_{\beta}$}

\newcommand{\Msun}{M$_{\odot}$}

\bibpunct[,]{(}{)}{;}{a}{,}{,}

\begin{document}

\title[PopStar Models II: Optical emission-line spectra from H{\sc ii} 
regions]{PopStar Evolutionary Synthesis Models II:
Optical emission-line spectra from Giant H{\sc ii} regions }

\author[Mart{\'{\i}}n-Manj{\'{o}}n et al.]
{M.L. Mart\'{\i}n-Manj\'{o}n$^{1}$
\thanks{E-mail:mariluz.martin@uam.es}, M.L.Garc\'{\i}a-Vargas$^{2,3}$,
M.~Moll\'{a}, $^{4}$\footnote{On sabatical leave on Institute of Astronomy, Department of Physics, University of Sydney, 2006 NSW, Australia} 
and A.~I.~D\'{\i}az$^{1}$\\ $^{1}$ Departamento de
F\'{\i}sica Te\'{o}rica, Universidad Aut\'onoma de
Madrid. Cantoblanco. E-28049 Madrid. Spain.\\ $^{2}$ FRACTAL
SLNE. C / Tulip\'{a}n 2, p13 1-A. E-28231 Las Rozas
(Madrid). Spain.\\ $^{3}$ Instituto de Astrof\'{\i}sica de Canarias.
C / V\'{\i}a L\'{a}ctea S/N E-38003 La Laguna (Tenerife). Spain.\\ $^{4}$
Departamento de Investigaci\'{o}n B\'{a}sica, CIEMAT,
Avda. Complutense 22. E-28040 Madrid. Spain.}

\date{Accepted Received ; in original form }

\pagerange{\pageref{firstpage}--\pageref{lastpage}} \pubyear{2008}

\maketitle

\begin{abstract}

This is the second paper of a series reporting the results from the
PopStar evolutionary synthesis models. Here we present synthetic
emission line spectra of H{\sc ii} regions photoionized by young star
clusters, for seven values of cluster masses and for ages between 0.1
and 5.2 Myr.  The ionizing Spectral Energy Distributions (SEDs) are
those obtained by the PopStar code \citep*{mgb09} for six different
metallicities, with a very low metallicity set, Z=0.0001, not included
in previous similar works.  We assume that the radius of the H{\sc ii}
region is the distance at which the ionized gas is deposited by the
action of the mechanical energy of the winds and supernovae from the
central ionizing young cluster. In this way the ionization parameter
is eliminated as free argument, since now its value is obtained from
the cluster physical properties (mass, age and metallicity) and from
the gaseous medium characteristics (density and abundances).  We
discuss our results and compare them with those from previous models
and also with a large and data set of giant H{\sc ii} regions for
which abundances have been derived in a homogeneous manner. The values
of the [OIII] lines (at $\lambda\lambda$ 4363, 4959, 5007\AA) in the
lowest metallicity nebulae are found to be very weak and similar to
those coming from very high metallicity regions (solar or
over-solar). Thus, the sole use of the oxygen lines is not enough to
distinguish between very low and very high metallicity regions. In
these cases we emphasize the need of the additional support of
alternative metallicity tracers, like the [SIII] lines in the
near-\textit{IR}.

\end{abstract}


\section{Introduction}

H{\sc ii} regions have been widely studied for the last three decades
\citep{pag80,eva85,dop86,diaz87,vil88,diaz91,diaz94}.  In those
studies the functional parameters, such as ionization parameter,
effective temperature of the ionizing stars and oxygen abundance, were
translated into physical parameters associated to the star clusters
(mass, age and metallicity).  In order to do this, several grids using
evolutionary models plus photoionization codes were computed by our
group \citep*{gd94,gbd95a,gbd95b,gmb98} and others
\citep{stas78,sta80,sta81,sta82,sta90}.

In the last decade the evolutionary synthesis technique has been
improved with the inclusion of updated stellar tracks or isochrones,
the use of better stellar spectra as input for the codes, a better
treatment of binary stars, a wider wavelength coverage and a higher
spectral resolution. Codes like {\sc starburst99}
\citep[hereinafter {\sc stb99}][]{lei99}, use the Geneva group
stellar tracks and are tuned for star forming regions. On the contrary
codes like {\sc pegase} \citep{fioc97} or {\sc galaxev} \citep{bru03}
are specially computed for intermediate and evolved populations and
use the Padova group stellar tracks. These codes, among others, have
been intensively applied to spectrophotometric catalogues to derive the
physical properties of stellar populations.

The results of these codes, mainly the SEDs, have been used in turn by
many authors to compute models for star forming regions, H{\sc ii} and
starburt galaxies, or have been applied to particular cases.
Recently, a more realistic approach to the H{\sc ii} region structure,
and therefore the ionization bubble, has been considered to improve
photoionization model results. Thus, \cite{moy01} analyzed coherently
the stellar and nebular energy distributions of starbursts and H{\sc
ii} galaxies, by using {\sc pegase} in addition to the photoionization
code {\sc cloudy} \citep*{fer98}, constructing models in which the
filling factor and the radius of the nebula are fixed to obtain a
range of values of the ionization parameter similar to those derived
from observations.  \cite{dop00} used both {\sc pegase} and {\sc
stb99} models to compute the SEDs of young star clusters.  Main
discrepances between the SEDs of these two grids of models were driven
by the different evolutionary tracks used, Padova's for {\sc pegase}
and Geneva's for {\sc stb99}.
They studied the emission line sequence and computed the H{\sc
ii} region spectra as a function of age, metallicity
and ionization parameter using the photoionization code {\sc mappings} v0.3
\citep{suth93,dop02,gro04}. More recently \cite{dop06a} used {\sc
stb99} plus {\sc mappings} III to produce a self-consistent model. In
this model they consider that the expansion and internal pressure of
the H{\sc ii} regions depends on the mechanical energy from the
central cluster, replacing the ionization parameter by a more
realistic one dependent on cluster mass and pressure in the ISM. A
similar approach has been used by \cite{stas03}. In this work they
calculate photoionization models of evolving starbursts by using as
ionizing continuum the SEDs from the code by \cite{sha98}. These
models were computed for different metallicity bins to compare with
the corresponding metallicity observations, assuming the geometry of
an adiabatic expanding bubble, instead of that of a region with a
fixed radius \citep[like in][]{gbd95a,gbd95b,stas96}.

The present work is the second paper of a series of three dedicated to
the PopStar models description and initial test-cases application.
PopStar is a new grid of evolutionary synthesis models described in
\citet*[][hereinafter Paper I]{mgb09}, where their suitability to model
stellar populations in a wide range of ages and metallicities is
shown.  These models are an updated version of \citet{gbd95b},
\citet*{gmb98} and \citet*{mgv00}.  One of the main improvements of
this new grid is a very careful treatment of the emerging spectrua of
hot stars, which can be either massive stars or post-AGB objects. In
previous works by our group \citep{gd94,gbd95a,gbd95b} hot massive
stars were modelled using the atmospheres of \cite{clegg87}. In the
new PopStar SEDs, NLTE blanketed models for O,B and WR stars by
\citet*{snc02} have been used. These models include a detailed
treatment of stellar winds which modifies the final SEDs obtained for
young stellar populations and consequently the emission line spectrum
of the surrounding nebula.

Our aim in the present work is to compute a set of updated
photoionization models, and therefore emission line spectra, for 
H{\sc ii} regions following the evolution of the ionizing young star
cluster, whose SED is computed by the evolutionary synthesis models
PopStar. We also compare these results with a sample of HII
regions where the metallicity has been carefully derived through the use of
appropriate calibrators. 

This grid of models has been computed without taking into account the
presence of dust and this imposes some limitations to its use. Our
purpose is to compute a grid of models covering the range of physical
parameters found in GEHR where the dust effect is weak on the optical
spectral lines, which are the ones computed in this work.  However, in
starburt galaxies, where star formation is taking place inside
molecular clouds where the dust is present, its effect can be very
important. The dust will absorb a fraction of the emitted ionizing
photons and therefore will affect the derived value the young cluster
mass using dust-free models.  The dust re-emits the light in the mid
and far infrared spectral regions, with evident consequences for the
population synthesis in these spectral windows
\citep*{bgs98,sil98,pan03,bre06,cle09b} and even in the radio
wavelength range \citep{veg08}.  Our purpose is to compute in the near
future self-consistent models including the chemical and the
spectro-photometric evolution, for spiral and irregular galaxies,
where star formation and dust effects are important, (since star
formation takes place inside molecular clouds).  In this case, the
molecular clouds will be specifically treated as a separated phase,
the enrichment of the gas will be well followed and the dust will be
included consistently.

Finally, we must mention that one of the main differences between the
isochrones from Geneva and Padova groups is the inclusion of stellar
rotation.  Stellar rotation affects the evolutionary tracks,
lifetimes, and chemical compositions of massive stars, as well as the
formation of red super-giants and WR stars \citep{mae08,mey05}. In
particular, the rotation of a star plays a determinant role at very
low metallicities, producing high mass loss where almost none was
expected, as explained by \citet[][ and references
therein]{eks08}. The revised grid of stellar evolutionary tracks
accounting for rotation, recently released by the Geneva group, has
been implemented into the {\sc stb99} evolutionary synthesis code
in the preliminary models by \citet{vaz07}.  Massive stars are
predicted to be hotter and more luminous than previously thought, this
effect being higher for decreasing metallicity. Individual stars now
tend to be bluer and more luminous, increasing by a factor of 2 (or
even more) the light-to-mass ratios at ultraviolet to near-infrared
wavelengths, as well as the total number of ionizing photons. However,
we have not taken this effect into account because we have used SEDs
based in a set of revised isochrones from the Padova group. This
revision affects mainly the computation of intermediate age stellar
populations (RGB and ABB phases), and does not include rotation.

Section 2 gives a summary of our theoretical evolutionary synthesis
and photoionization models, the inputs and the hypotheses assumed for
the grid calculation. Section 3 shows our model results in terms of
the time evolution of emission lines. In this section we also include
the comparison with the previous models by \cite{gbd95a, gbd95b}. In
Section 4 we present a discussion of the models based on the
comparison with observed emission line ratios which have been
carefully compiled from the literature for objects whose abundances
have been consistently re-calculated using the appropriate
techniques. This section also includes the guidelines for deriving the
physical properties of the ionizing star clusters through the analysis
of the correponding gas emission line spectra. Finally, our
conclusions are given in Section 5.

\section{Summary of theoretical models}
\subsection{Evolutionary synthesis models.}

The SEDs used in this work have been taken from the new PopStar
evolutionary models (Paper I). The basic model grid is composed by
Single Stellar Populations (hereinafter SSPs) for 6 different Initial
Mass Functions (IMF) from which only those with a Salpeter power law
\citep{sal55} and different mass limits have been used here:
0.85\Msun\ and 120M$_{\odot}$ (hereinafter SAL1), 0.15\Msun\ and
100M$_{\odot}$ (hereinafter SAL2), and 1\Msun\ and 100M$_{\odot}$ (
hereinafter STB). None of the models include either binaries or mass
segregation.

Isochrones are an update of those from \cite{bgs98} for 6 different
metallicities: 0.0001. 0.0004, 0.004, 0.008, 0.02 and 0.05 
the lowest metallicity set being now added. 
The age coverage is from $\log{t}=5.00$ to 10.30 with a
variable time resolution of $\Delta(\log{t})=0.01$ in the youngest
stellar ages. The details of the isochrones are described in Paper I.

Atmosphere models are from \cite{lcb97} with an excellent coverage in
effective temperature, gravity and metallicities, for stars with Teff
$\leq 25000$K.  For O, B and WR the code uses the NLTE blanketed
models by \cite{snc02} at $\rm Z = 0.001$, 0.004, 0.008, 0.02 and
0.04.  There are no models available for Z = 0.0001, 0.0004 and
Z=0.04. For these stars, we have selected the models with the closest
metallicity, that is, Z=0.04 for the isochrones with Z=0.05 and
Z=0.001 for isochrones with Z=0.0001 and 0.0004. The possible
misselection for the lowest metallicities is not important for what
refers to WR stars since, in principle, there are no WR at these
lowest Z. This is in any case a limitation of the models: the grid for
NLTE atmosphere models is not as fine as the LTE Lejeune.  There are
110 models for O-B stars, calculated by \cite{phl01}, with 25000K $<$
Teff $\leq 51500$K and $2.95 \leq \log{g} \leq 4.00$, and 120 models
for WR stars (60 WN and 60 WC), from \cite{hm98}, with 30000K $\leq
T^{*} \leq 120000$K and $1.3R_{\odot}\leq R^{*}\leq 20.3 R_{\odot}$
for WN, and with $ 40000K \leq T^{*} \leq 140000$K and $
0.8R_{\odot}\leq R^{*}\leq 9.3 R_{\odot}$ for WC. T$^{*}$ and R$^{*}$
are the temperature and the radius at a Roseland optical depth of
10. The assignation of the appropriate WR model is consistently made
by using the relationships among opacity, mass loss and velocity wind
described in Paper I.

For post-AGB and PN with Teff higher than 50000K and up to 220000K the
NLTE models by \cite{rau03} are taken. For higher temperatures PopStar
uses black bodies. The use of these last models affects the resulting
intermediate age SEDs, which are not used in the present work. 

\subsection{Photoionization models for H{\sc ii} regions.}

We have studied the evolution of a cluster along the first 5.2 Myr, in
21 steps of time. Clusters older than this age do not have the
necessary ionization photons to produce a visible emission line
spectrum, although they are still detectable on \halpha\
images\footnote{See Garc\'{\i}a-Vargas, Moll\'{a} \&
Mart\'{\i}n-Manj\'{o}n, in preparation, where we calculate the
photometric properties of young star clusters taking into account the
contamination by the emission lines and the underlying old populations
where these clusters are embedded in the computed colors, for an
analysis about the cluster physical properties determined from the
photometric information only}. Ionizing clusters have been assumed to
form in a single burst with different masses. We have run models
assuming seven values for the total cluster mass, 0.12, 0.20, 0.40,
0.60, 1.00, 1.50 and 2.00 $\times 10^{5}$ M$_{\odot}$, using SAL2 IMF
(with m$_{low}$=0.15\Msun\ and m$_{up}$=100 M$_{\odot}$).  Masses in
this range are able to provide the observed number of ionizing photons
of most medium to large extragalactic H{\sc ii} regions. Also, in
order to compare with our old photoionization models we have used SAL1
IMF.

We have used the photoionization code {\sc CLOUDY} \citep{fer98} to
obtain the emission line spectra of the modelled H{\sc ii} regions for
different metallicities. The gas is assumed to be ionized by the
massive stars of the young cluster whose ionizing spectra have been
taken fro the PopStar SEDs computed in Paper I, as explained in the
previous section.  The SEDs are given for a normalized mass of 1 \Msun, 
hence the stellar mass of the cluster must be used to scale the number of
ionizing photons and other absolute parameters given by PopStar for
each IMF.  The shape of the ionizing continuum, the number of ionizing
photons, Q(H), and the H{\sc ii} region radius, R$_{s}$, set by the
action of the cluster mechanical energy, are obtained directly from
the ionizing cluster parameters: mass, age and metallicity.

The chemical composition of the gas and its spatial distribution,
assuming a given geometry, together with the medium density, are used
as inputs for the photoionization code. It is assumed that both the
cluster and the surrounding gas have the same chemical
composition. The solar abundances are taken from \cite{gre98}.  The
use of these solar abundances implies that $Z_{\odot}=0.017$
\footnote{Asplund, Grevesse \& Sauval (2005) obtained a solar abundance
$Z_{\odot}=0.012$. This value is still questioned \citep*{bah05,dra05,ant05} 
because it does not fit the helioseismological constraints.}, 
thus, Z $=$ 0.02 of our models does
not correspond to the solar value but to 1.17 Z$_{\odot}$.

The solar abundances are summarized in Table~\ref{abun}, Column 2. 
 Some refractary elements, Na, Al, Si, Ca, Fe and Ni, must be depleted
due to the presence of dust grains mixed with the ionized gas which
can affect the \textit{UV} photon absorption and decrease the electronic
temperature.
\begin{scriptsize}
\begin{table}
\caption{Abundances, $log{X/H}$ used in the models \citep{gre98}}.
\begin{tabular}{lrrrrrrrr}
\hline
El. & sol. & sol. & 0.0001 & 0.0004 & 0.004& 0.008& 0.02& 0.05 \\
        &       & depl&depl&depl&depl&depl&depl&depl \\
\hline
He &-1.09 &-1.09& -1.24& -1.24 &-1.20 &-1.16 &-1.07& -0.90 \\
C  &-3.48 &-3.48& -5.71& -5.11 &-4.11 &-3.81 &-3.41& -3.01 \\
N  &-4.08 &-3.96& -6.19& -5.59 &-4.59 &-4.29 &-3.89& -3.49 \\
O  &-3.17 &-3.18& -5.41& -4.81 &-3.81 &-3.51 &-3.11& -2.71 \\
Ne &-3.92 &-3.92& -6.15& -5.55 &-4.55 &-4.25 &-3.85& -3.45 \\
Na &-5.67 &-6.67& -8.90& -8.30 &-7.30 &-7.00 &-6.60& -6.20 \\
Mg &-4.42 &-5.42& -7.65& -7.05 &-6.05 &-5.75 &-5.35& -4.95 \\
Al &-5.53 &-6.53& -8.76& -8.16 &-7.16 &-6.86 &-6.46& -6.06 \\
Si &-4.45 &-4.75& -6.98& -6.38 &-5.38 &-5.08 &-4.68& -4.28 \\
S  &-4.67 &-4.67& -6.90& -6.30 &-5.30 &-5.00 &-4.60& -4.20 \\
Ar &-5.60 &-5.60& -7.83& -7.23 &-6.23 &-5.93 &-5.53& -5.13 \\
Ca &-5.64 &-6.64& -8.87& -8.27 &-7.27 &-6.97 &-6.57& -6.17 \\
Fe &-4.50 &-5.50& -7.73& -7.13 &-6.13 &-5.83 &-5.43& -5.03 \\
Ni &-5.75 &-6.75& -8.98& -8.38 &-7.38 &-7.08 &-6.68& -6.28 \\
\hline
\label{abun}
\end{tabular}
\end{table}
\end{scriptsize}
 Solar depleted abundances, obtained using
\citet*{gar95} depletion factors, are in Column 3. For each
metallicity, the element abundances heavier than helium have been
scaled by a constant factor, with respect to the hydrogen content,
according to the solar depleted abundances, as given in Columns 4 to 8
of the table. 

The radius of the modelled region has been derived from the mechanical energy
produced by the expanding atmospheres of massive stars with strong
winds.  \citet*{cmw75} demonstrated that an early-type star with a
strong stellar wind can blow out a large cavity or  {\sl bubble} in the
surrounding gas. The wind-driven shell begins to evolve with an
initial phase of free expansion followed by an adiabatic expansion
phase, and then the material shall collapse into a thin, cold shell as a
result of radiative cooling. At this stage the gas shall trap the
ionization front and thus the radiative phase begins. In this phase the
ionizing photons are absorbed and the region cools via emission in the
Balmer lines. In this process, the radius of the outer shock, R$_{s}$,
evolves as:
\begin{equation}
R_{s}=1.6(\epsilon/n)^{1/5} t^{3/5} (pc)
\end{equation}
\onecolumn
\footnotesize
\begin{table*}
\caption{Emission line spectrum from a H{\sc ii} region of mass
4x10$^{4}$ \Msun\ and Z=0.008 as a function of cluster parameters for
n$_{H} =$ 10 cm$^{-3}$. This is Table 2a. Table 2b includes the models
with n$_{H} =$ 100 cm$^{-3}$. Both tables for the whole set of
cluster masses and metallicities are provided in electronic format.}
\begin{tabular}{cccccccccccccc}
\hline
Zmet & $log{age}$ &  $Mass$    & $[OII]$  & $[OIII]$ & $[OIII]$ & $[OIII]$ &  $[OI]$  &  $[NII]$ & $[NII]$  &  $[SII]$    &  \\
     & yr  & M$_{\odot}$ & 3727\AA\ & 5007\AA\ & 4959\AA\ & 4363\AA\ & 6300\AA\ & 6584\AA\ & 6548\AA\ &6716\AA\   \\
 & & & & & & & & & & & & & \\
 \hline
 & & & & & & & & & & & & & \\
0.008&5.00&4$ 10^{4}$&0.7398&3.5490&1.1791&0.0140&0.0189&0.2101&0.0712&0.1413&\\
0.008&5.48&4$ 10^{4}$&0.7644&3.8260&1.2711&0.0162&0.0206&0.2121&0.0719&0.1492&\\
0.008&5.70&4$ 10^{4}$&0.7802&3.9562&1.3144&0.0173&0.0214&0.2147&0.0728&0.1535&\\
0.008&5.85&4$ 10^{4}$&0.8049&4.1260&1.3708&0.0186&0.0222&0.2189&0.0742&0.1585&\\
0.008&6.00&4$ 10^{4}$&0.8196&3.8744&1.2872&0.0168&0.0221&0.2246&0.0761&0.1592&\\
0.008&6.10&4$ 10^{4}$&0.8279&3.6580&1.2153&0.0154&0.0217&0.2282&0.0773&0.1583&\\
0.008&6.18&4$ 10^{4}$&0.8788&3.5548&1.1810&0.0147&0.0221&0.2386&0.0809&0.1638&\\
0.008&6.24&4$ 10^{4}$&1.0206&3.3794&1.1227&0.0135&0.0211&0.2476&0.0839&0.1671&\\
0.008&6.30&4$ 10^{4}$&1.2825&2.9609&0.9837&0.0111&0.0201&0.2722&0.0923&0.1745&\\
0.008&6.35&4$ 10^{4}$&1.4038&2.4940&0.8286&0.0086&0.0198&0.3065&0.1039&0.1884&\\
0.008&6.40&4$ 10^{4}$&1.6003&1.8690&0.6209&0.0057&0.0182&0.3542&0.1200&0.2073&\\
0.008&6.44&4$ 10^{4}$&1.6950&1.6437&0.5461&0.0048&0.0183&0.3994&0.1354&0.2296&\\
0.008&6.48&4$ 10^{4}$&1.7781&1.2606&0.4188&0.0032&0.0164&0.4705&0.1594&0.2616&\\
0.008&6.51&4$ 10^{4}$&1.4910&1.5379&0.5109&0.0042&0.0221&0.4548&0.1541&0.2791&\\
0.008&6.54&4$ 10^{4}$&1.8240&2.2962&0.7629&0.0082&0.0459&0.5305&0.1798&0.3893&\\
0.008&6.57&4$ 10^{4}$&2.2596&2.5285&0.8400&0.0099&0.0712&0.6845&0.2320&0.5287&\\
0.008&6.60&4$ 10^{4}$&2.8121&2.1852&0.7260&0.0083&0.1159&0.9505&0.3221&0.7827&\\
0.008&6.63&4$ 10^{4}$&2.2239&1.3071&0.4342&0.0037&0.0838&0.8983&0.3044&0.7102&\\
0.008&6.65&4$ 10^{4}$&2.2603&0.8811&0.2927&0.0022&0.0790&0.9855&0.3340&0.7794&\\
0.008&6.68&4$ 10^{4}$&2.4881&0.4758&0.1581&0.0011&0.0813&1.1627&0.3940&0.9343&\\
0.008&6.70&4$ 10^{4}$&2.4949&0.2760&0.0917&0.0000&0.0776&1.2372&0.4192&1.0078&\\
0.008&6.72&4$ 10^{4}$&2.3095&0.1077&0.0358&0.0000&0.0718&1.2473&0.4227&1.0720&\\
\hline

$[SII]$ & $[SIII]$ & $[SIII]$ & $[SIII]$ & $[NeII]$ & $He\,I$ & $He\,I$ & $He\,II$ & 
$H_{\alpha}$ & log L$H_{\beta}$ & $\log{u}$ & R$_{s}$ \\
 6731\AA\ &  6312\AA\ & 9069\AA\ & 9532\AA\ & 3869\AA\ & 4471\AA\ &  5836\AA\ & 4686\AA\ &  6563\AA\ & ergs$^{-1}$ &  &  pc \\
 & & & & & & & & & & & & \\
 \hline
  & & & & & & & & & & & & \\
0.0995&0.0140&0.2966&0.7357&0.2600&0.0336&0.0843&0.0000&2.9635&38.727&-0.56&11.00\\
0.1050&0.0149&0.3013&0.7472&0.2848&0.0335&0.0841&0.0000&2.9606&38.734&-1.03&19.00\\
0.1081&0.0154&0.3051&0.7567&0.2967&0.0335&0.0840&0.0000&2.9590&38.741&-1.21&23.64\\
0.1117&0.0161&0.3128&0.7756&0.3115&0.0335&0.0839&0.0000&2.9566&38.753&-1.37&28.69\\
0.1121&0.0157&0.3170&0.7862&0.2931&0.0335&0.0840&0.0000&2.9605&38.747&-1.54&34.70\\
0.1115&0.0152&0.3189&0.7908&0.2768&0.0335&0.0841&0.0000&2.9633&38.754&-1.64&39.23\\
0.1153&0.0154&0.3293&0.8167&0.2702&0.0336&0.0841&0.0000&2.9634&38.765&-1.77&46.18\\
0.1175&0.0161&0.3511&0.8707&0.2545&0.0335&0.0841&0.0000&2.9624&38.761&-1.85&50.84\\
0.1227&0.0164&0.3696&0.9166&0.2130&0.0335&0.0840&0.0000&2.9627&38.750&-1.98&57.81\\
0.1323&0.0152&0.3601&0.8930&0.1725&0.0335&0.0841&0.0000&2.9673&38.732&-2.08&63.93\\
0.1455&0.0138&0.3450&0.8557&0.0991&0.0335&0.0841&0.0000&2.9727&38.674&-2.17&66.13\\
0.1610&0.0132&0.3385&0.8394&0.0846&0.0334&0.0840&0.0000&2.9732&38.651&-2.25&70.45\\
0.1833&0.0119&0.3214&0.7971&0.0593&0.0333&0.0835&0.0000&2.9752&38.602&-2.32&72.37\\
0.1955&0.0110&0.3021&0.7492&0.1035&0.0335&0.0841&0.0000&2.9825&38.599&-2.37&76.71\\
0.2732&0.0147&0.3374&0.8366&0.2293&0.0335&0.0841&0.0027&2.9705&38.575&-2.55&91.74\\
0.3712&0.0166&0.3526&0.8743&0.2968&0.0334&0.0836&0.0110&2.9631&38.727&-2.72&104.4\\
0.5492&0.0159&0.3313&0.8215&0.3348&0.0330&0.0823&0.0333&2.9637&38.518&-2.96&118.2\\
0.4971&0.0102&0.2604&0.6459&0.1979&0.0332&0.0833&0.0135&2.9785&38.388&-2.97&105.1\\
0.5449&0.0083&0.2271&0.5633&0.1422&0.0332&0.0833&0.0047&2.9784&38.275&-3.07&109.2\\
0.6526&0.0069&0.1918&0.4756&0.0845&0.0326&0.0816&0.0023&2.9726&38.205&-3.25&115.2\\
0.7034&0.0057&0.1637&0.4059&0.0554&0.0315&0.0788&0.0027&2.9710&38.074&-3.35&120.3\\
0.7474&0.0039&0.1201&0.2980&0.0242&0.0285&0.0713&0.0000&2.9720&38.009&-3.50&125.8\\
\hline
\end{tabular}
\label{table2}
\end{table*}
\twocolumn
\normalsize

where $\epsilon$ is the total injected mechanical energy (SN and
stellar winds) per unit time in units of 10$^{36}$ ergs s$^{-1}$, $n$
is the interstellar medium density in units of cm$^{-3}$, and $t$ is
the age of the shell in units of 10$^{4}$ yr. Since we use the
value of the energy at each time-step, this radius represents the
instantaneous size of the region obtained by adding the energy
produced by stellar winds and SN from one time step to the
next. Therefore $\epsilon$ is this energy divided by the time
step. We have used the radius given for each time $t$ and 1~\Msun\ in
Table 4 from Paper I, scaled to the stellar cluster mass of each model
extrapolating this bubble geometry to a shell structure formed by the
combined effects of the mechanical energy deposition from the massive
stars winds and SN explosions belonging to the ionizing cluster. The
ionized gas is assumed to be located in a thin spherical shell at
distance R$_{s}$ from the ionizing source. This approach has the
advantage of eliminating the ionization parameter as a free parameter
in the models, being now computed from the physical properties of the
evolving young cluster. This radius is still large enough when
compared to the shell thickness as to validate the approximation of a
plane-parallel geometry.

 The hydrogen density has been considered constant throughout the
nebula and an ionization-bounded geometry has been assumed, with the
hydrogen density equal to the electron density for complete ionization
(Case B of H recombination). We have computed models with two
different values of n$_{H}$: 10~cm$^{-3}$ and 100~cm$^{-3}$, in order
to check the density effect on the emitted spectrum. A density of 10
cm$^{-3}$ is representative of small-medium isolated H{\sc ii} regions
\citep{cas02a, pm05} while 100 cm$^{-3}$ is more appropriate for
modelling H{\sc ii} galaxies \citep{hag08} and large circumnuclear
H{\sc ii} regions \citep{gv97, diaz07}, frecuently found around the
nuclei of starbursts and AGNs. Although the constant density
hypothesis is probably not realistic, it can be considered
representative when the integrated spectrum of the nebula is analyzed.

Using the number of ionizing photons, Q(H), the nebula size, R$_{s}$,
and the hydrogen density, n$_{H}$ we can obtain the ionization
parameter, u, given by:
\begin{equation}
u = \frac{Q(H)}{4 \pi c n_{H} R_{s}^{2}}
\end{equation}
where $c$ is the speed of light. Under these assumptions, the parameter
$u$ recovers its physical meaning since it is directly derived from
the physical cluster's properties. Hence, for a given IMF and n$_{H}$,
each model is computed as a function of the cluster mass, age and
metallicity.  The values of $u$ derived for our set of models are 
within the range of observed values for H{\sc ii} regions.

Since the current set of models have been computed for medium to large
extragalactic H{\sc ii} regions, whose measured reddening is usually
very low and often consistent with the galactic reddening, we have not
included dust effects, as explained before, except for the abundance
depletion.  Only when these models are used for obscured compact H{\sc
ii} regions or for large star forming regions like starbursts
galaxies, dust effects should be taken into account.

\section{Results}

\subsection{Emission lines.}

Once the photoionization code is applied, we obtain the emission line
spectra of the associated H{\sc ii} region produced by the ionizing
cluster for every set of models. We have followed the evolution of the
emitted spectrum during 5.2 Myr after the cluster formation. After
this time, no emission line spectrum is produced due to the lack of
ionizing photons. In the present study we have considered only the
most intense emission lines in the optical spectrum which dominate the
cooling for moderate metallicities. In higher (over solar) metallicity
regions the cooling is shifted to the infrared.

The models results, assuming a SAL2 IMF for all masses and
metallicities, are summarized in Table~\ref{table2} a) and b), for
values of n$_{H}$ o 10~cm$^{-3}$ and 100~cm$^{-3}$
respectively. Listed in columns 1 to 3 are: Z (metallicity), logarithm
of the age in years, cluster mass (in \Msun); in columns 4 to 20 we
give the intensities relative to \hbeta\ of the following lines:
[OII]$\lambda$3727, [OIII]$\lambda$5007, [OIII]$\lambda$4959,
[OIII]$\lambda$4363, [OI]$\lambda$6300, [NII]$\lambda$6584,
[NII]$\lambda$6548, [SII]$\lambda$6716, [SII]$\lambda$6731,
[SIII]$\lambda$6312, [SIII]$\lambda$9069, [SIII]$\lambda$9532,
[NeIII]$\lambda$3869, He\,I$\lambda$4471, He\,I$\lambda$5836,
He\,II$\lambda$4686, H$_{\alpha}$; finally, in columns 21 to 23 we
list the logarithm of the intensity of H$_{\beta}$ in erg$\cdot$
s$^{-1}$ ($log LH\beta$), the logarithm of the ionization parameter
($\log{u}$) and the ionized region radius in pc ($R_{s}$).

Part of Table~\ref{table2}a for Z$=$0.008 and a cluster mass of
4$\times$10$^{4}$ \Msun is shown as an example. The complete tables
are available in electronic format.

\subsection{Comparison with previous models}

\begin{figure*}
\resizebox{0.70\hsize}{!}{\includegraphics[clip,angle=0]{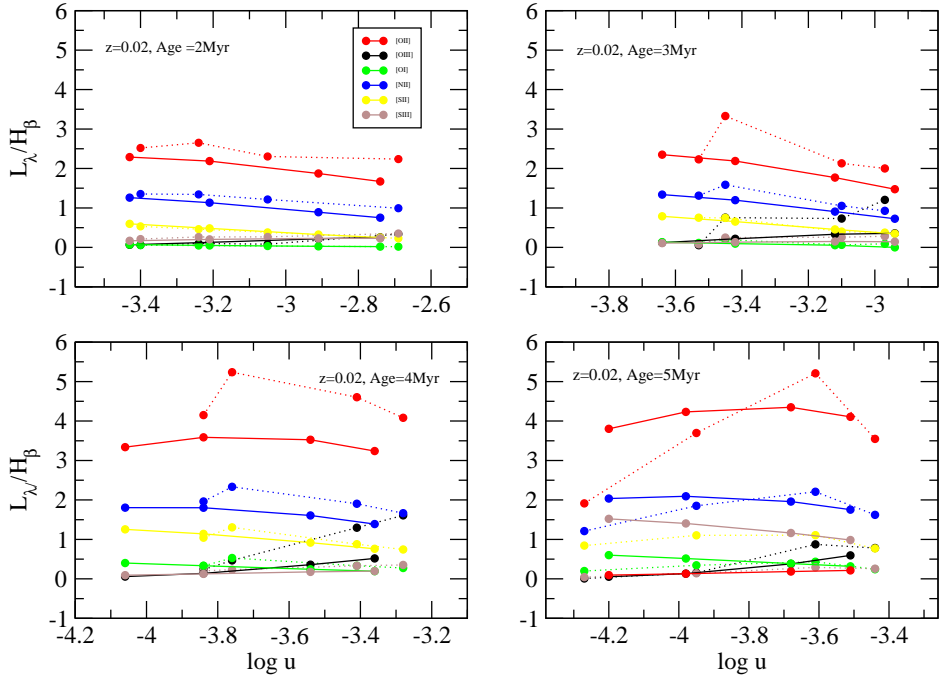}}
\caption{Evolution of emission lines at Z $= 0.02$ and SAL1 IMF for $\rm n_{H} =
10 cm^{-3}$ and $\log{R} (cm) = 20.84$.  Dotted lines correspond to old 
models by \citet{gbd95b} while solid lines correspond to new models.}
\label{gv95_z02}
\end{figure*}

\begin{figure*}
\resizebox{0.70\hsize}{!}{\includegraphics[angle=0]{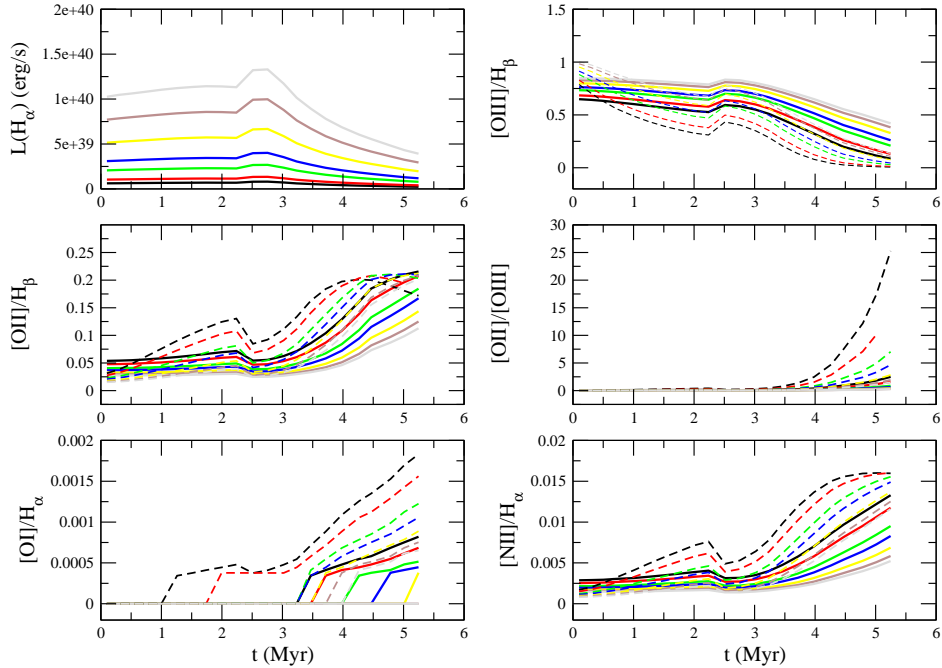}}
\caption{a). Evolution of the \halpha\ luminosity and five emission line
ratios for Z $=$ 0.0001 (SAL2). The ratios are
[OIII]$\lambda\lambda$5007,4959/H$_{\beta}$,
[OII]$\lambda$3727/H$_{\beta}$, [OII]/[OIII],
[OI]$\lambda$6300/H$_{\alpha}$ and
[NII]$\lambda$6584/H$_{\alpha}$. The more appropriate hydrogen
recombination line has been chosen for normalization in order to
minimize reddening effects. Solid lines correspond to models with
n$_{H} =$~10 cm$^{-3}$ while dotted lines correspond to models with
n$_{H} =$~100 cm$^{-3}$. Black, red, green, blue, yellow, brown and
grey lines correspond to different cluster masses: 0.12, 0.20, 0.40,
0.60, 1.00, 1.50 and 2.00 $\times$ 10$^{5}$~M\sun, respectively. }
\setcounter{figure}{1}
\label{evol_lines}
\end{figure*}

In order to evaluate the effect of the changes incorporated into the
new models we have compared our present results with the previous ones
given by \citet{gbd95a} by computing a set of models using the same
input abundances and the same assumptions for the ionization parameter
as in \cite{gbd95b}. In those models the ionization parameter was
calculated with a fixed radius of $\rm \log{R_{s}}(cm) = 20.84$. The
results are shown in Fig.~\ref{gv95_z02}, where the emission line
ratios as a function of the ionization parameter $\log{u}$, for Z $=$
0.02 at ages of 2 Myr, 3 Myr, 4 Myr and 5 Myr can be seen. We have
plotted the results for the following emission lines:
[OII]$\lambda$3727, [OIII]$\lambda$5007, [OI]$\lambda$6300,
[NII]$\lambda$6584, [SII]$\lambda$6731 and [SIII]$\lambda$9069. In
these figures, the dotted lines correspond to the models from
\cite{gbd95b} while the solid lines correspond to the models presented
here.

We can see that, in general, the emission lines resulting from
\cite{gbd95b} models are more intense than in the models computed in
the present work for the same value of the ionization parameter. This
is due to the use in the new set of models of NLTE blanketed
atmosphere models for massive stars, which produce fewer hard ionizing
photons. This new set of models explains in a natural way the emission
line ratios found in low excitation high metallicity H{\sc ii}
regions, as we will show in Section 4 and therefore do not require
{\sl ad hoc} explanations to keep the number of hard ionizing photons
to the observed values.

\subsection{Emission-line evolution}

In this work we follow the evolution of a given cluster during 5.2 Myr
from their formation. After this time, in most cases and depending on
metallicity, the gaseous emission lines are too weak to be measured
due to the paucity of ionizing photons.

Figs \ref{evol_lines}a) to \ref{evol_lines}e) show the time evolution
of the H$_{\alpha}$ luminosity and several emission line ratios, as
labelled, selected to represent the most relevant ionization stages of
the most common elements: [OIII]$\lambda\lambda$5007,4959/H$_{\beta}$,
[OII]$\lambda$3727/H$_{\beta}$,
[OII]$\lambda$3727/[OIII]$\lambda\lambda$5007,4959,
[OI]$\lambda$6300/H$_{\alpha}$ and [NII]$\lambda$6584/H$_{\alpha}$ for
models of different metallicity, computed with SAL2 IMF.  In each
figure, different colours correspond to a different cluster masses,
from 0.12~$\times$~10$^{5}$ to 2~$\times$ 10$^{5}$~M$_{\odot}$. The
two different density cases, n$_{H}$=10~cm$^{-3}$ and
n$_{H}$=100~cm$^{-3}$, are represented by solid and dashed lines,
respectively.

\begin{figure*}
\resizebox{0.70\hsize}{!}{\includegraphics[angle=0]{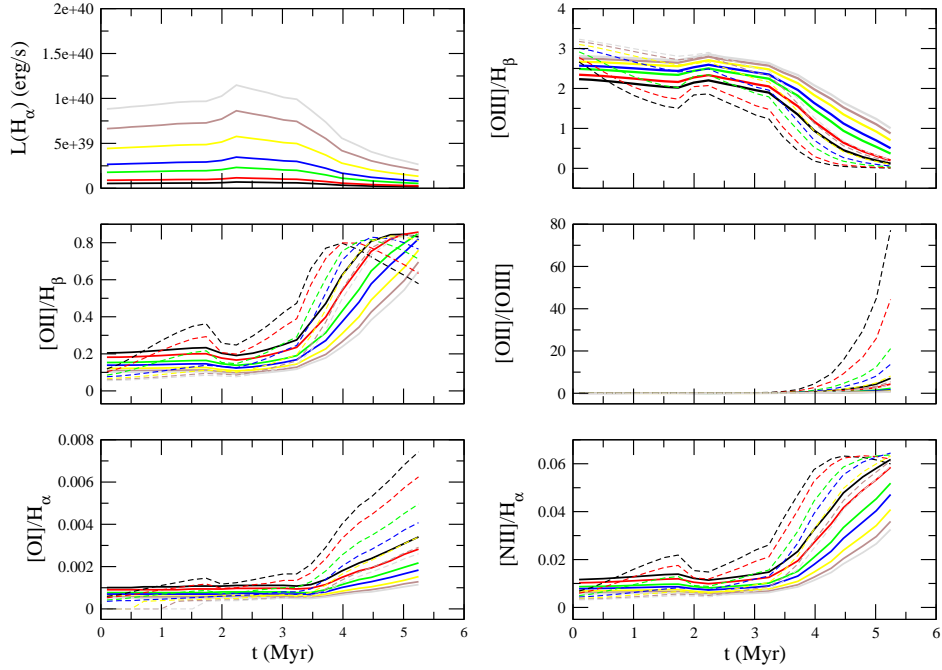}}
\caption{b).Evolution of emission lines for Z $=$ 0.0004 (SAL2). Different
colors and line code have the same meaning as in Fig.~\ref{evol_lines}a.}
\setcounter{figure}{1}
\end{figure*}

\begin{figure*}
\resizebox{0.70\hsize}{!}{\includegraphics[angle=0]{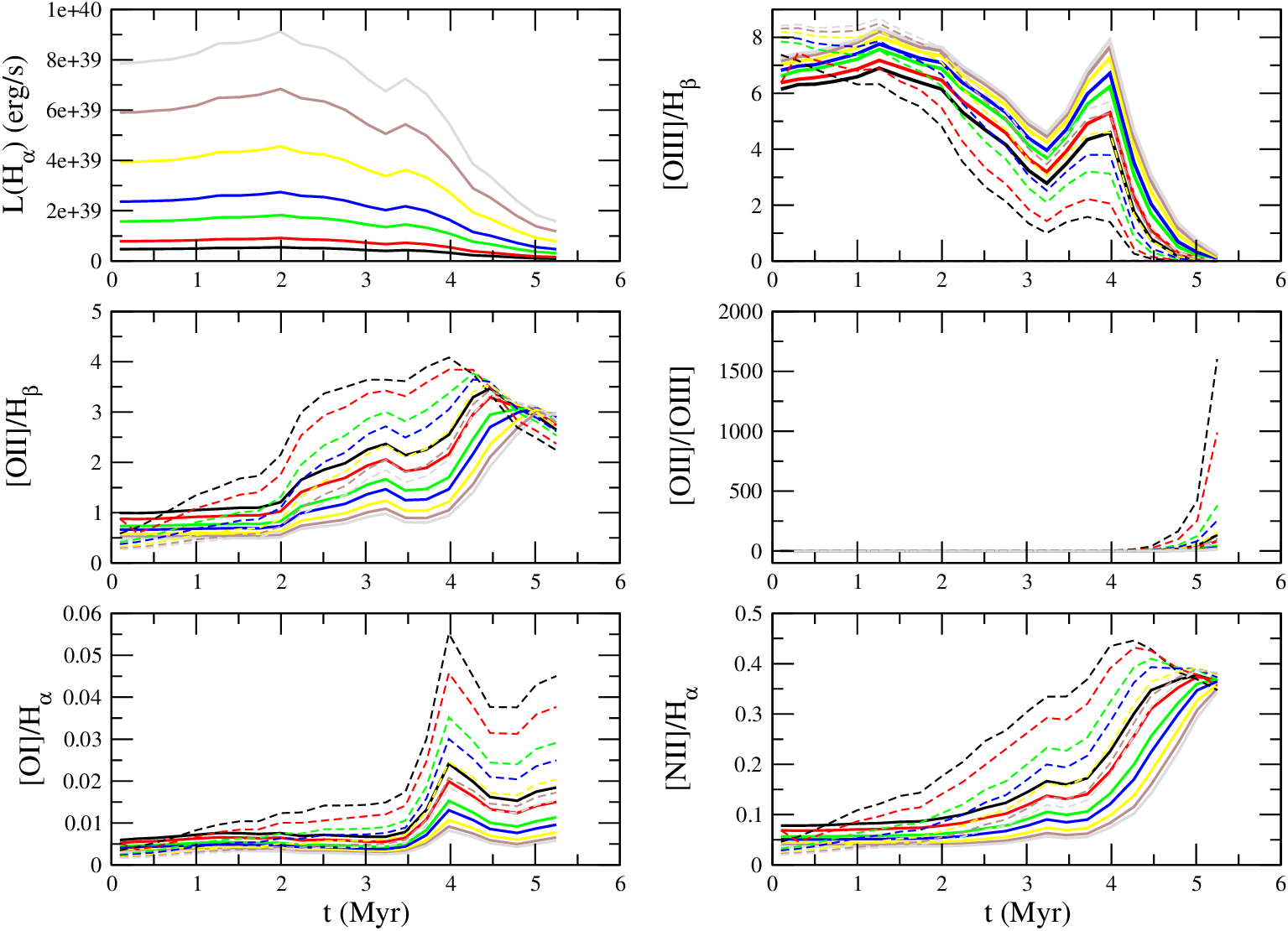}}
\caption{c). Evolution of emission lines for Z $=$ 0.004 (SAL2). Different
colors and line code have the same meaning as in Fig.~\ref{evol_lines}a.}
\setcounter{figure}{1}
\end{figure*}

\begin{figure*}
\resizebox{0.70\hsize}{!}{\includegraphics[angle=0]{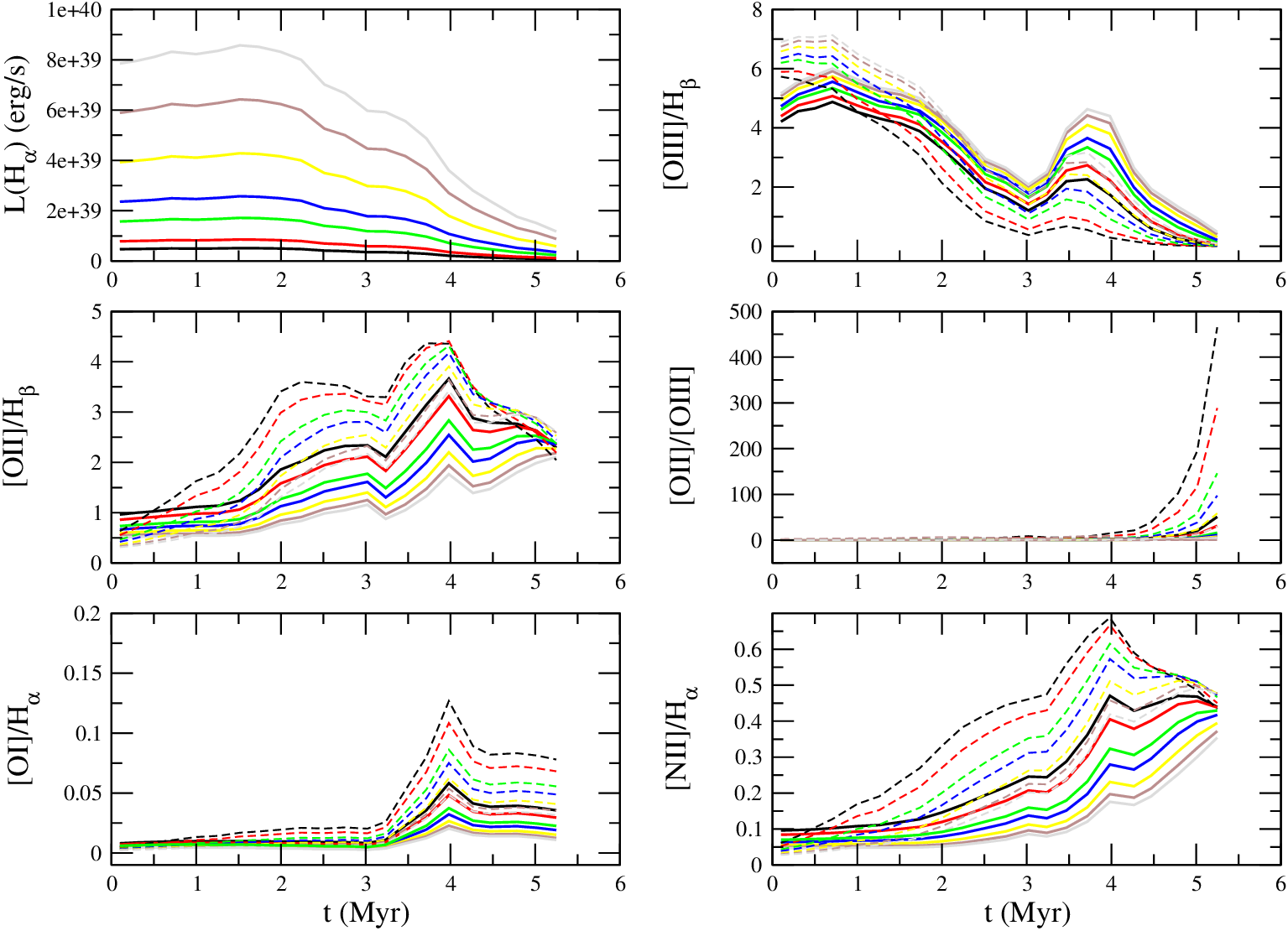}}
\caption{d). Evolution of emission lines for Z $=$ 0.008 (SAL2). Different
colors and line code have the same meaning as in Fig.~\ref{evol_lines}a.}
\setcounter{figure}{1}
\end{figure*}

\begin{figure*}
\resizebox{0.70\hsize}{!}{\includegraphics[angle=0]{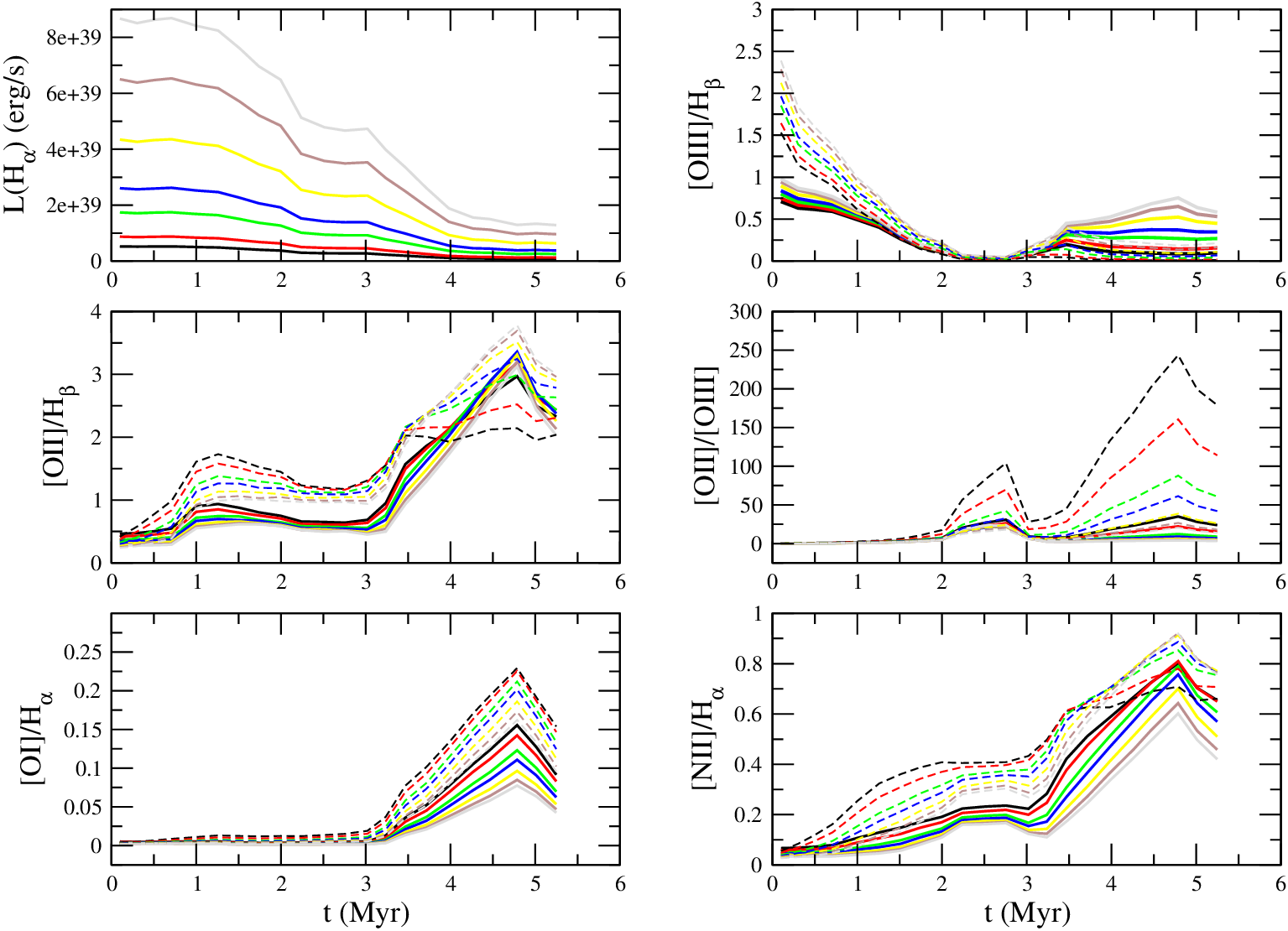}}
\caption{e). Evolution of emission lines for Z $=$ 0.02 (SAL2). Different
colors and line code have the same meaning as in Fig.~\ref{evol_lines}a.}
\end{figure*}

For the lowest abundance models (Z$<$ 0.0001) emission line ratios
involving oxygen lines are very weak (note the different scale in each
metallicity panel) and its evolution is smooth. For intermediate
metallicity cases, the emission lines are intense, increasing with
cluster mass. 
\begin{figure}
\resizebox{0.90\hsize}{!}{\includegraphics[clip,angle=0]{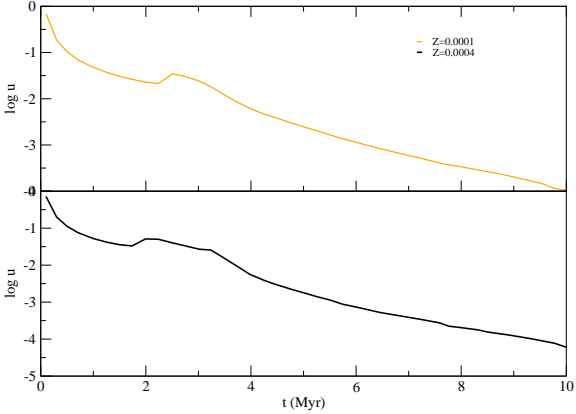}}
\caption{The ionization parameter as a function of age (beyond 
5.2 Myr) for a 1 $\times$ 10$^{5}$ \Msun\ cluster, low metallicities (Z $=$ 0.0001 and Z $=$ 0.0004), and n$_{H} =$~10 cm$^{-3}$} 
\label{uextended}
\end{figure}

\begin{figure}
\resizebox{0.90\hsize}{!}{\includegraphics[clip,angle=0]{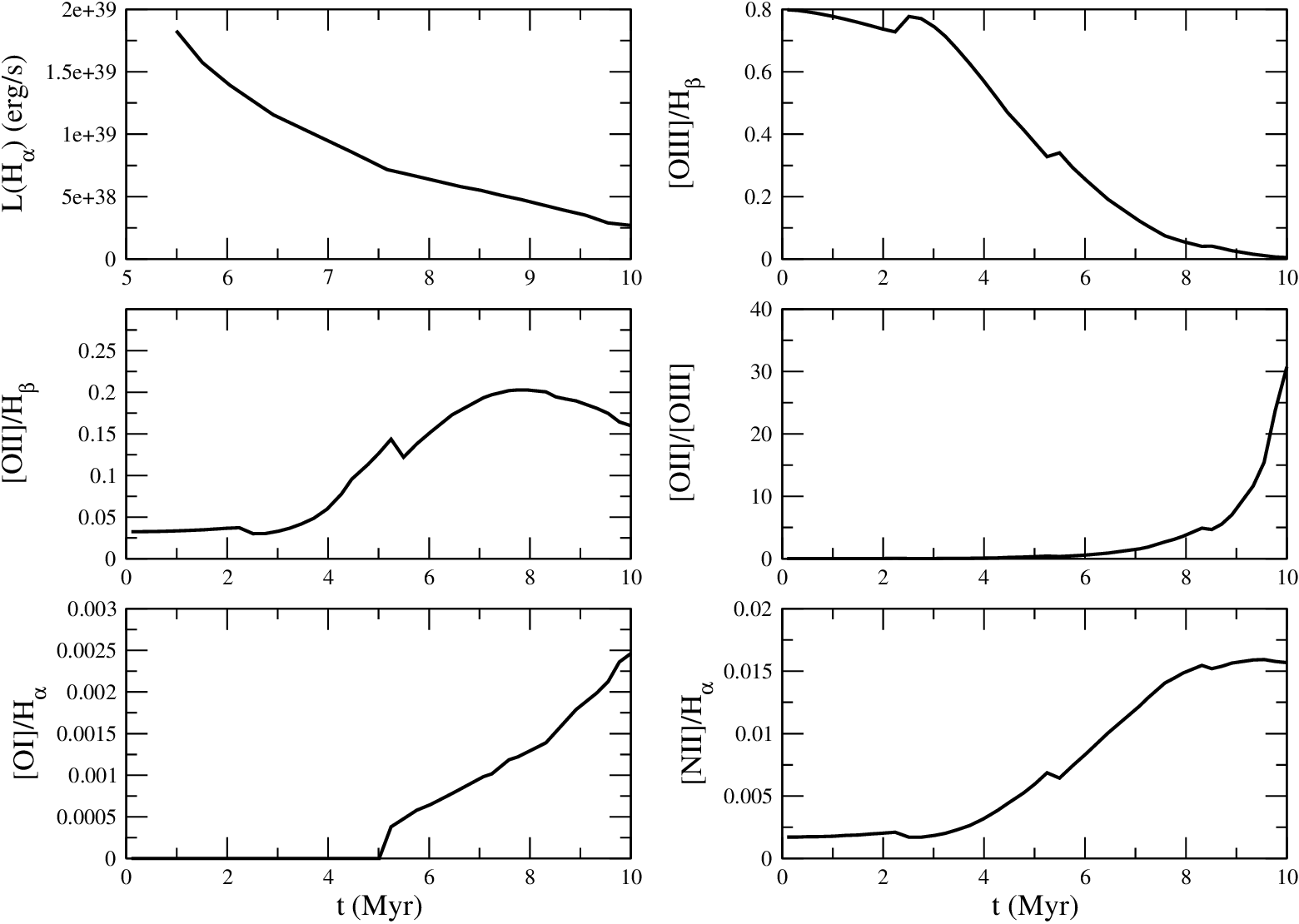}}
\caption{Extendended evolution (uo to 10 Myr) of the H$_{\alpha}$
luminosity and five emission line ratios:
[OIII]]$\lambda\lambda$5007,4959/H$_{\beta}$,
[OII]$\lambda$3727/H$_{\beta}$, [OII]/[OIII],
[OI]$\lambda$6300/H$_{\alpha}$ and [NII]$\lambda$6584/H$_{\alpha}$,
for a 1 $\times$ 10$^{5}$ \Msun\ cluster, with Z=0.0001 (SAL2) and
n$_{H} =$~10 cm$^{-3}$.}
\label{evol_lines_extendedz0001}
\end{figure}

\begin{figure}
\resizebox{0.90\hsize}{!}{\includegraphics[clip, angle=0]{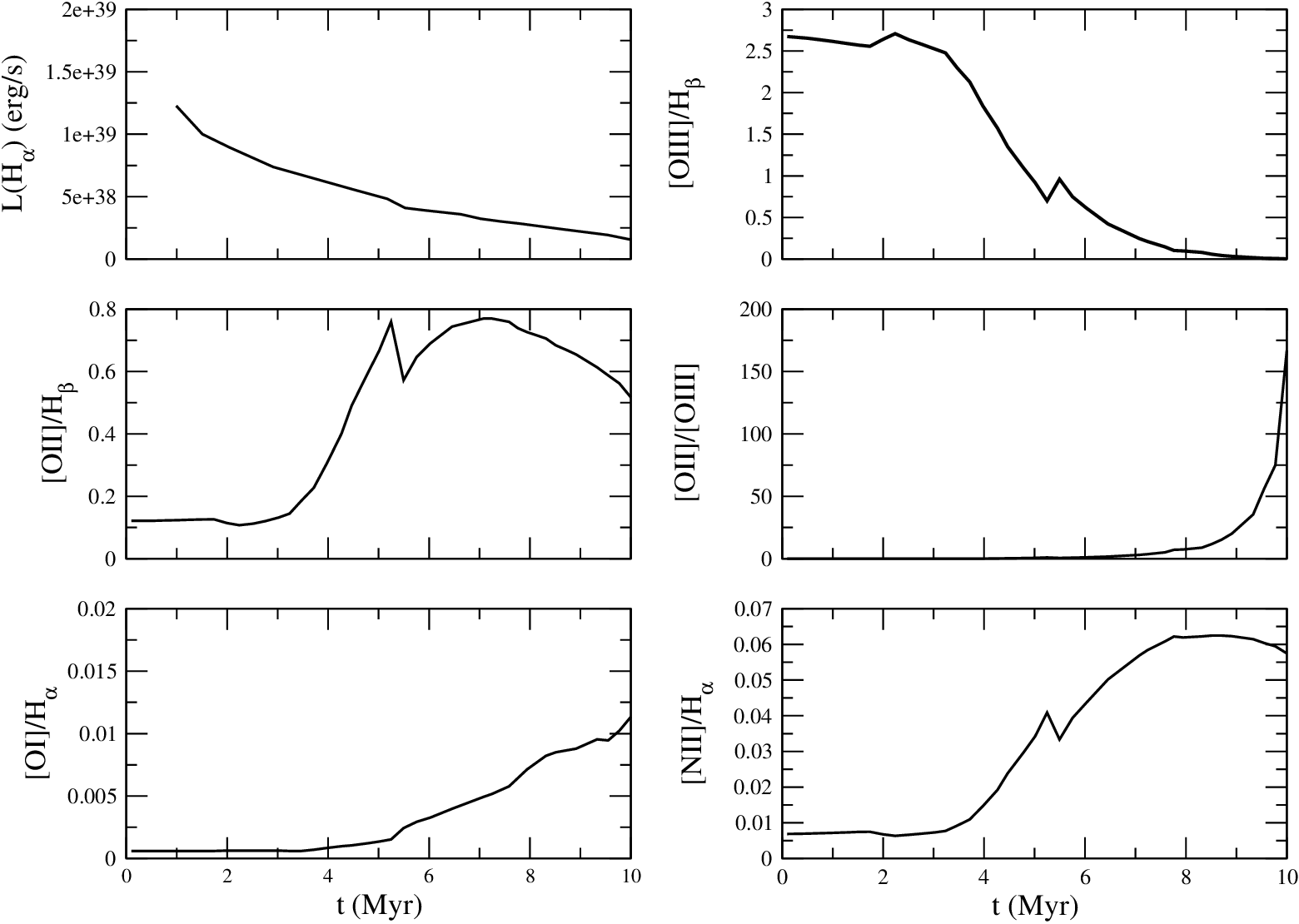}}
\caption{Extendended evolution (uo to 10 Myr) of the H$_{\alpha}$
luminosity and five emission line ratios:
[OIII]]$\lambda\lambda$5007,4959/H$_{\beta}$,
[OII]$\lambda$3727/H$_{\beta}$, [OII]/[OIII],
[OI]$\lambda$6300/H$_{\alpha}$ and [NII]$\lambda$6584/H$_{\alpha}$,
for a 1 $\times$ 10$^{5}$ \Msun\ cluster, with Z=0.0004 (SAL2) and
n$_{H} =$~10 cm$^{-3}$.}
\label{evol_lines_extendedz0004}
\end{figure}

In the first Myr of the evolution, the
[OIII]]$\lambda\lambda$5007,4959 lines are intense and then they
decrease to rise again at 3-4 Myr, due to the presence of WR stars,
that makes the equivalent effective temperature of the cluster become
higher and causes an increase of the excitation during a short
time. The appearance of the WR features and their duration depends on
the mass and metallicity of the cluster as discussed in Paper I.  The
[OII]$\lambda$3727 line shows moderate variations with
metallicity. The [OII]/H$_{\beta}$ ratio increases after 3 Myr of the
star formation, being almost constant for low metallicity models,
before this age. As a result, the [OII]/[OIII] ratio is very low for
all metallicities at ages younger than 4 Myr.  The oxygen emission
lines are most intense for models with Z=0.004. We will come back to
the oxygen line evolution in the next section, when studying the
diagnostic diagrams.

The low ionization line [OI]$\lambda$6300 is an indicator of the
presence of strong winds and supernovae \citep{stas96}, and therefore,
a burst age indicator.  On the sole basis on the photo-ionization
models, the intensity of this line increases with cluster age and
[O/H] abundance, therefore it shows higher values in old, evolved and
high metallicity H{\sc ii} regions. The addition of shock
contributions (not included in our models), would render this value
even higher.

Finally, the [NII]$\lambda$6584/H$\alpha$ line intensity ratio, which
is in principle a good metallicity indicator, increase with increasing
metallicity, even for regions with metallicities higher than Z=0.004.

We have also produced photoionization models for ages older than 5.2
Myr (until 10 Myr) for the lowest metallicity cases, Z=0.0001, 0.0004
(see Fig.~\ref{uextended}). The lower the metallicity, the hotter the
ZAMS and the longer the time spent by stars in the ZAMS. Hence, for
the lowest metallicity clusters (Z=0.0001, 0.0004) we have explored
the evolution further, to 10 Myr. However, the ionization parameters
obtained are low, producing very weak (and sometimes undetectable)
emission lines from high-ionization stages, like [OIII], specially for
the lowest masses. Of course, these regions do exist and produce a
low-ionization spectrum, with observable [OI] and [OII] and [NII]
lines but negligigle [OIII] line intensities which place them out of
the plotted area of the diagnostic diagrams used in this
work. Figures~\ref{evol_lines_extendedz0001}
and~\ref{evol_lines_extendedz0004} illustrate this fact showing the
results of the photoionization models for the 10 Myr evolution of a
ionizing cluster of 1 $\times$ 10$^{5}$\Msun\ and a density of n$_{H}
=$~10 cm$^{-3}$ for two metallicities: Z=0.0001 and Z=0.0004.

\begin{figure}
\resizebox{0.90\hsize}{!}{\includegraphics[clip,angle=0]{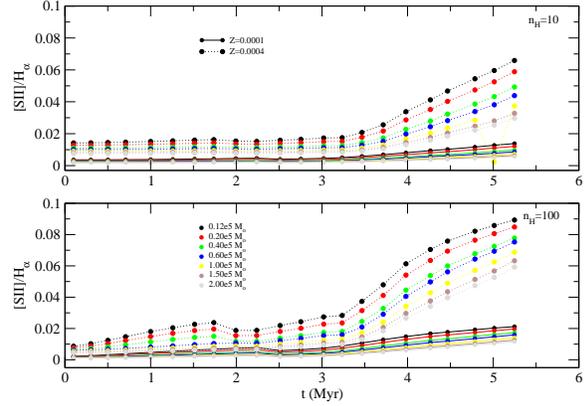}}
\caption{The [SII]$\lambda\lambda$6717,31/\halpha\ line ratio {\sl vs} age (in
Myr) for low metallicities (Z=0.0001 and Z=0.0004) and all masses.
Solid lines join models with Z=0.0001 while dotted lines join models
whith Z=0.0004. Different colors correspond to different values of the
cluster mass.  Models with values of n$_{H} =$~10 cm$^{-3}$ have been
plotted in the upper panel while models with n$_{H} =$~100 cm$^{-3}$
have been plotted in the lower one.}
\label{s2t}
\end{figure}

Figs.~\ref{s2t} and~\ref{s23t} show the resulting sulphur emission
line ratios: [SII]$\lambda\lambda$6717,31/\halpha\ and
[SII]$\lambda\lambda$6717,31/[SIII]$\lambda\lambda$9069,9532, as a
function of the burst age. Sulphur line ratios were proposed
as ionization parameter indicator by \citet{diaz91} and calibrations
based on photoionization models at different metallicities were
presented in \citet{diaz2000,diazmonterp2000?}.  We have carried out a
recalibration of this parameter using the models presented here that
is detailed in Appendix A. Since the ionization parameter is directly
related to the stellar cluster age, as suggested by \cite{gbd95a}, we
propose to use these sulphur lines as age indicators for H{\sc ii}
regions. Fig.~\ref{s2t} shows the
[SII]$\lambda\lambda$6717,31/\halpha\ ratio for the lowest
metallicities (Z=0.0001 and Z=0.0004) and all masses. Models with
different values of n$_{H}$ have been plotted in different panels. For
a given metallicity, the plot shows that the line ratios involving the
sulphur lines are low and rather constant during the first 3 Myr of
evolution. From 3 to 5 Myr, they increase smoothly with age for all
cluster masses. The line ratios also increase with cluster mass at a
given age but cluster mass can be constrained via the derived number
of ionizing photons from Balmer line luminosities.

\begin{figure}
\resizebox{0.90\hsize}{!}{\includegraphics[clip,angle=0]{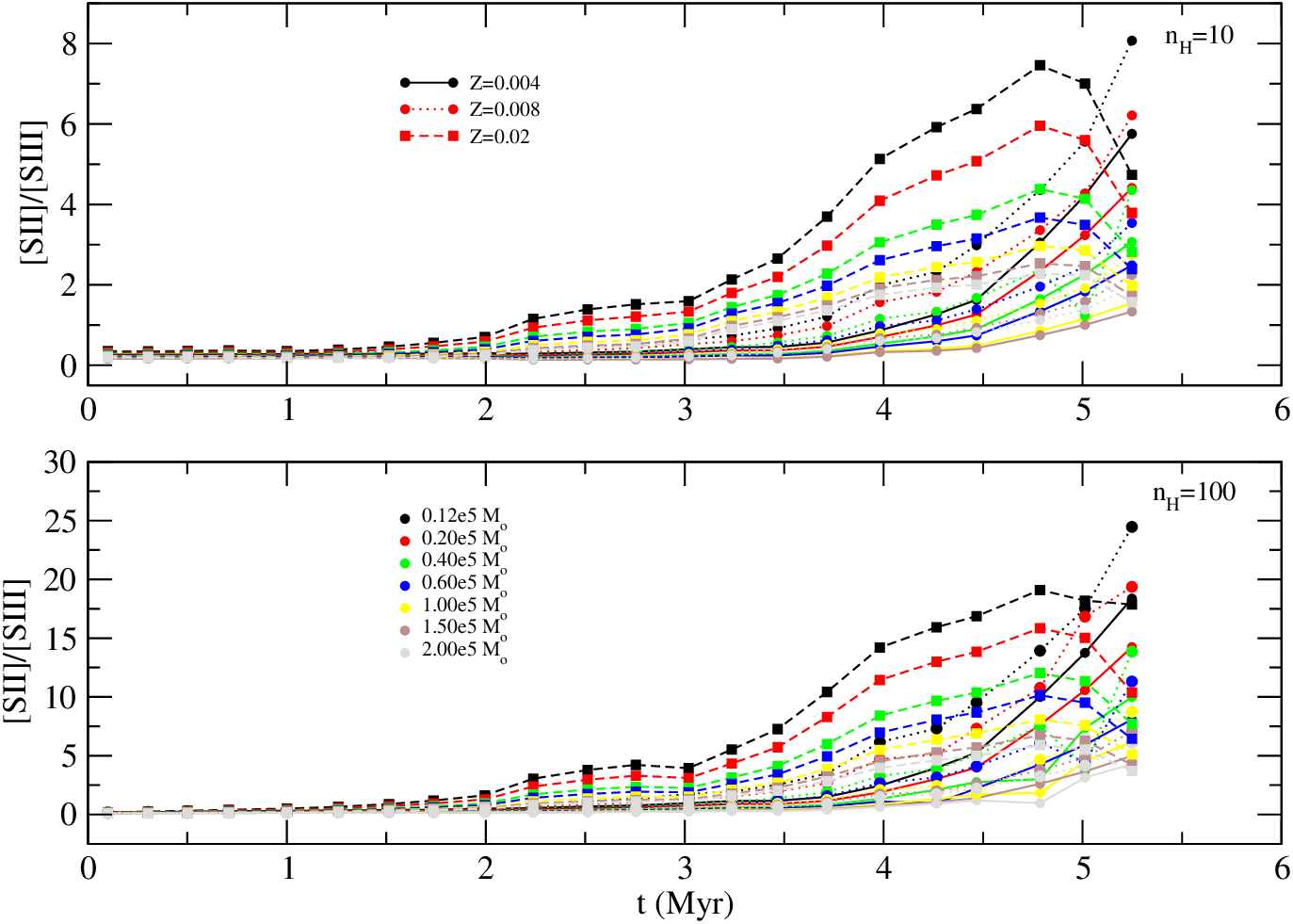}}
\caption{The
[SII]$\lambda\lambda$6717,31/[SIII]$\lambda\lambda$9069,9532 line
ratio {\sl vs} age (in Myr) for high metallicities (Z=0.004, 0.008,
0.02) and all masses.  Solid lines join models with Z=0.004, dotted
lines those with Z=0.008 and dashed lines those with Z=0.02. Different
colors correspond to different values of the cluster mass. Models with
values of n$_{H} =$~10 cm$^{-3}$ have been plotted in the upper panel
while models with n$_{H} =$~100 cm$^{-3}$ have been plotted in the
lower one.}
\label{s23t}
\end{figure}

\subsection{Emission line intensities and the ionization parameter. }

Figs.~\ref{parameter_u} a) to e) show the intensity, with
respect to \hbeta\, of the following emission lines:
[OII]$\lambda$3727, [OIII]$\lambda$5007 , [OI]$\lambda$6300,
[NII]$\lambda$6584, [SII]$\lambda$6731, [SIII]$\lambda$9069 {\sl vs}
$\log{u}$, obtained from the computed models, for metallicities Z =
0.0001, 0.0004, 0.004, 0.008 and 0.02.  In each figure four panels are
shown for ages of 2 Myr (top left), 3 Myr (top right), 4 Myr (bottom
left) and 5 Myr (bottom right).  Solid lines correspond to n$_{H}=$ 10
cm$^{-3}$ models while dotted lines correspond to n$_{H}=$ 100
cm$^{-3}$. In total 7 points per line are shown corresponding to 7
values of the cluster mass: 0.12, 0.20, 0.40, 0.60, 1.00, 1.50 and
2.00 $\times 10^{5}$ \Msun. We can see in the figures that models with
n$_{H}=$ 10 cm$^{-3}$ and 100 cm$^{-3}$ follow the same trend, and
cover different ranges of the ionization parameter value, with this
being higher, as expected, for the lower density case.

\begin{figure*}
\resizebox{0.70\hsize}{!}{\includegraphics[clip,angle=0]{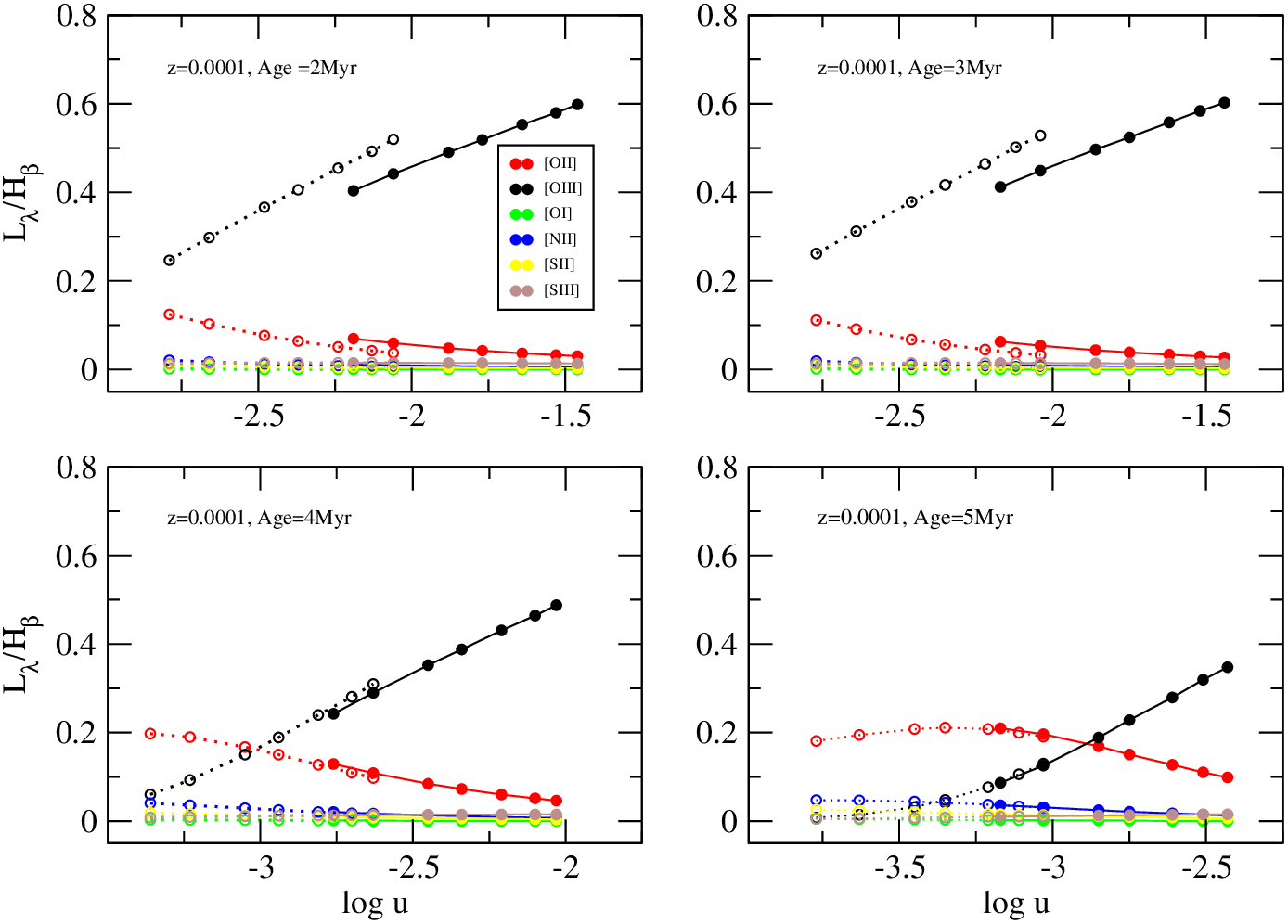}}
\caption{a). Evolution of emission line intensities, as labelled, normalized to the H$\beta$ intemsity for Z $=$ 0.0001 (SAL2) models as a
function of the logarithm of the ionization parameter. Solid lines
correspond to models with n$_{H} =$ 10 cm$^{-3}$ and dotted lines
correspond to models with n$_{H} =$ 100 cm$^{-3}$.}
\setcounter{figure}{7}
\label{parameter_u}
\end{figure*}

\begin{figure*}
\resizebox{0.70\hsize}{!}{\includegraphics[clip,angle=0]{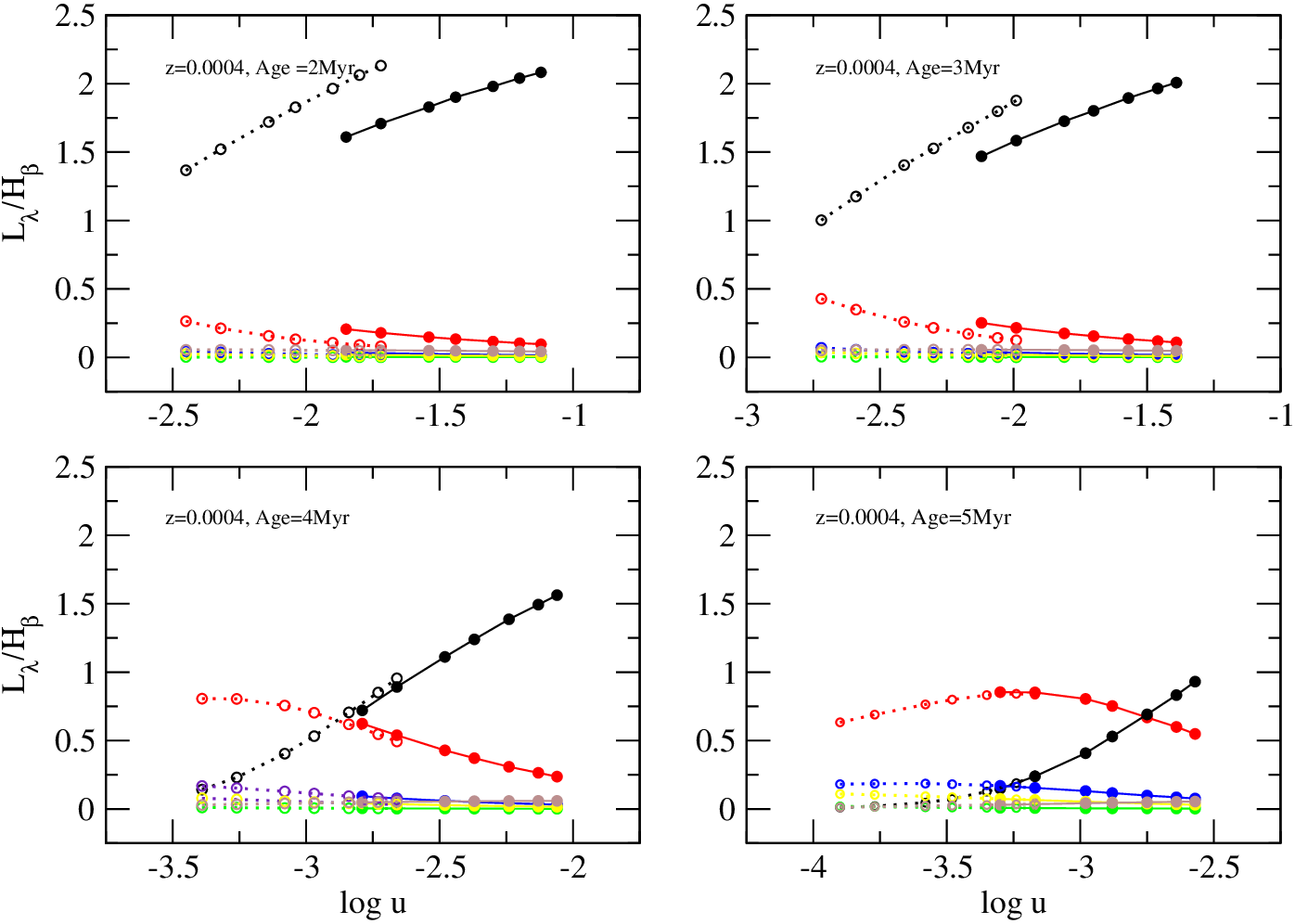}}
\caption{b). Same as Fig.~\ref{parameter_u} but for Z=0.0004.}
\setcounter{figure}{7}
\end{figure*}

\begin{figure*}
\resizebox{0.70\hsize}{!}{\includegraphics[clip,angle=0]{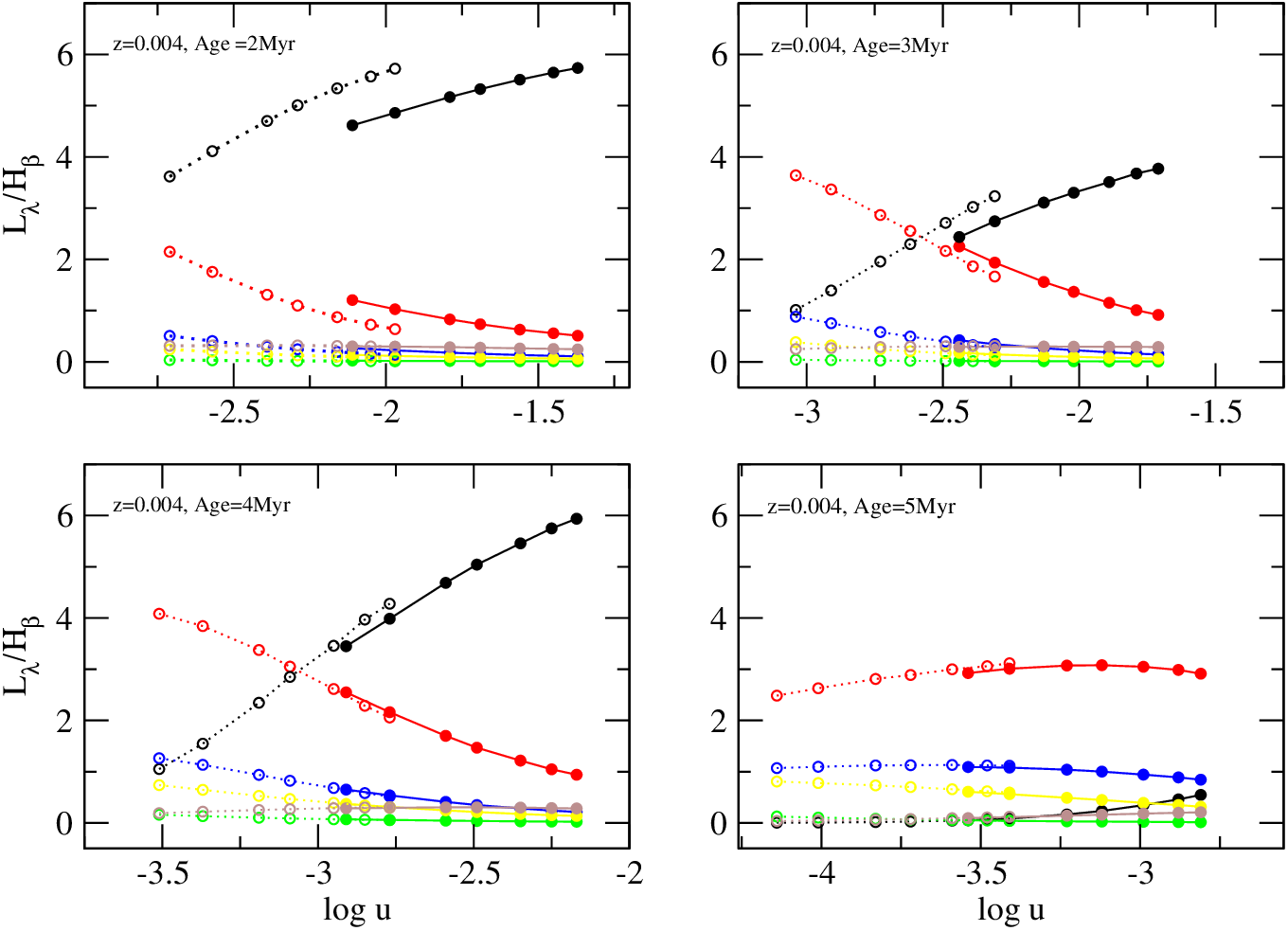}}
\caption{c). Same as Fig.~\ref{parameter_u} but for Z=0.004.}
\setcounter{figure}{7}
\end{figure*}

\begin{figure*}
\resizebox{0.70\hsize}{!}{\includegraphics[clip,angle=0]{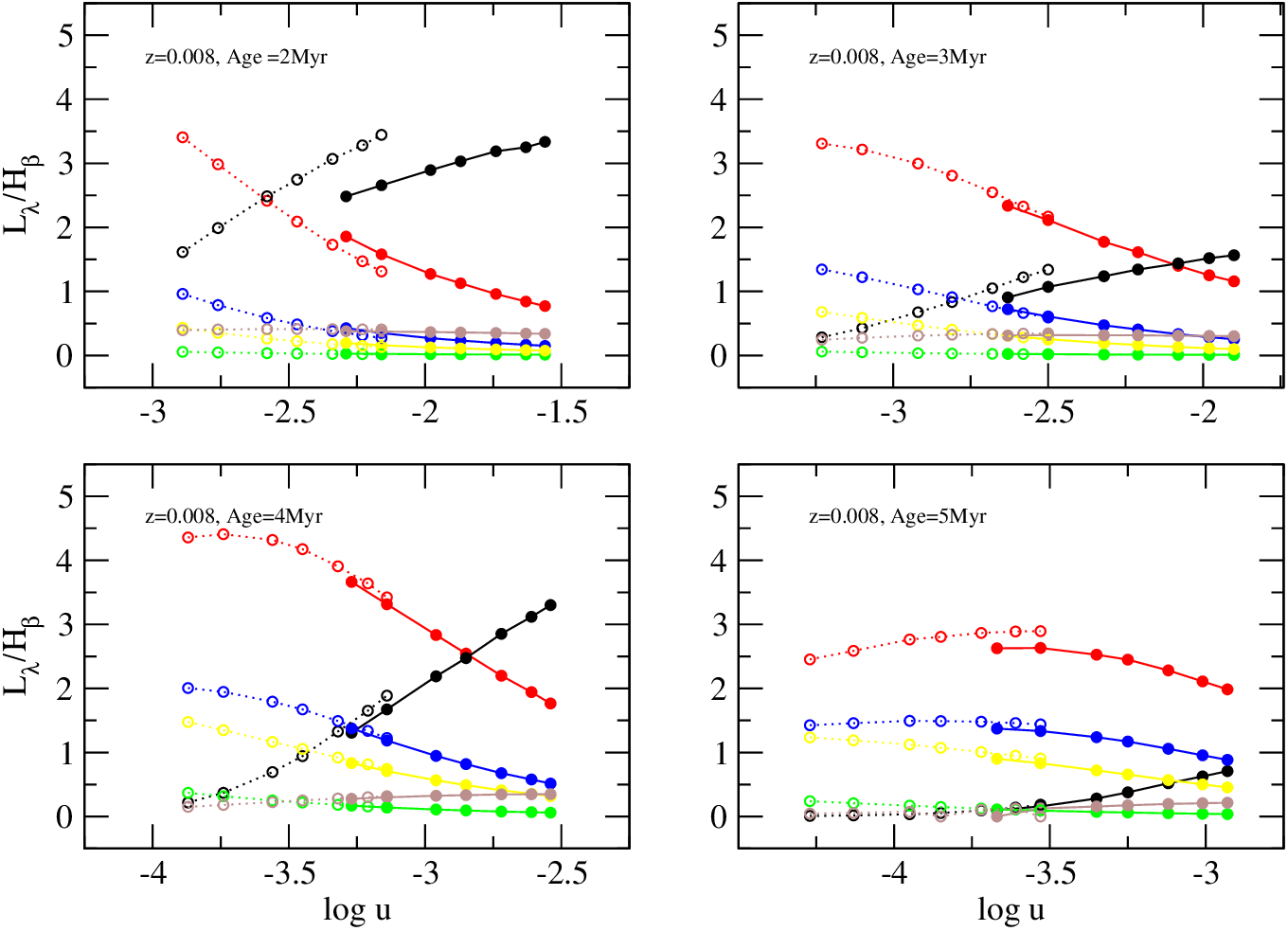}}
\caption{d). Same as Fig.~\ref{parameter_u} but for Z=0.008.}
\setcounter{figure}{7}
\end{figure*}

\begin{figure*}
\resizebox{0.70\hsize}{!}{\includegraphics[clip,angle=0]{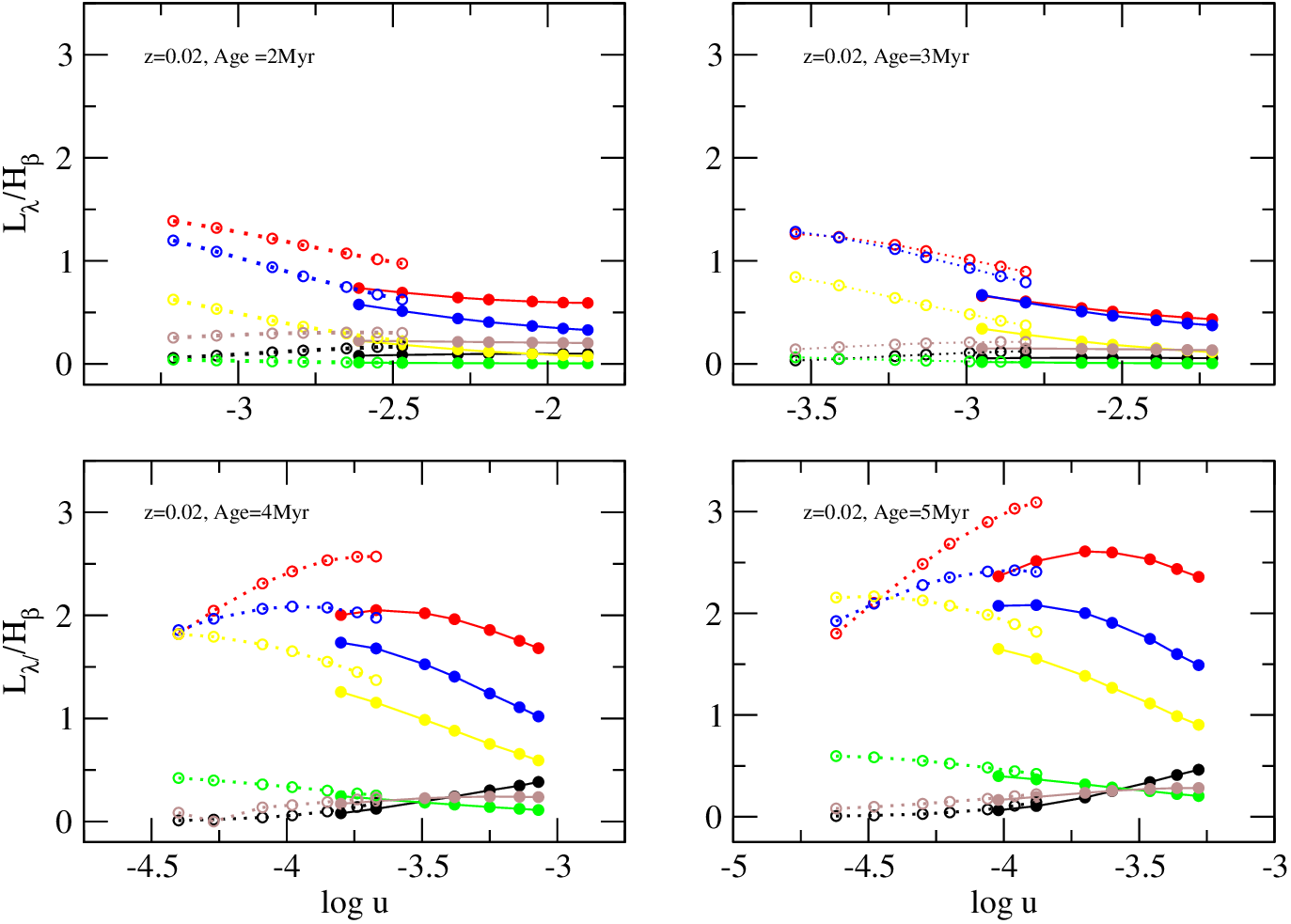}}
\caption{e). Same as Fig.~\ref{parameter_u} but for Z=0.02.}
\end{figure*}

For a given metallicity, the changes in the emission line spectrum and
in the ionization parameter are due to the changes in the cluster
mass, which determines the number of ionizing photons, and in the
cluster age, which influences not only the total number of ionizing
photons but also the overall ionization spectrum hardness and the
ionized region size. For example, the [OIII]$\lambda$5007 line is
intense during the first few Myr of the cluster evolution, for all
metallicities, and then it decreases, to rise again at 4 Myr due to
the presence of WR stars that produces a harder ionizing continuum.

From our models we predict an intrinsic sizing effect in the H{\sc ii}
region evolutionary sequence, which implies a natural decreasing of
the ionization parameter. For a very young region whose ionization is
dominated by the O-B stars, the size is still small but the number of
ionizing photons is high, resulting in a high ionization parameter. As
the cluster evolves, the region size increases due to the stellar
winds, but the number of ionizing photons decreases only slightly
producing a decreasing ionization parameter, and therefore lower
intensity lines.  For the highest metallicities, this effect is more
remarkable. High metallicity regions have less ionizing photons and
more intense winds therefore implying a lower ionization parameter. In
addition, the effective cooling of the gas is transferred from the
optical emission lines to the near and mid infrared ones as
metallicity increases. These two facts explain the fact that in high
metallicity regions, the optical emission lines are extremely weak,
reaching in most cases only about one per cent of the H$_{\beta}$ line
intensity. These lines are difficult to detect and this fact has
biased the observational samples during a long time towards regions of
a restricted metallicity range which has important implications for
the interpretation of photometric samples of H{\sc ii} regions.

\section{Discussion}

\subsection{Observational Data}

We have used first a compilation of data on H{\sc ii} galaxies, high
and low metallicity H{\sc ii} regions, and circumnuclear star forming
regions regions (CNSFR), with emission line intensities measured both
for the auroral and nebular [SIII] lines at $\lambda$ 6312 and
$\lambda\lambda$ 9069,9532 respectively. The data for CNSFR can be
found in \citet{diaz07} where the appropriate references for the rest
of the objects in the compilation can also be found.  For all the
regions, the electron density has been derived using the [SII] ratio
\citep{ost89}.

For this sample, metallicities have been derived following standard
techniques in the cases in which electron temperatures could be
derived. Otherwise, an empirical calibration based on the
S$_{23}$/O$_{23}$ parameter (see \citet{diaz07}) has been used.

Data from \citet{cas02a} correspond to H{\sc ii} regions in the spiral
galaxies: NGC628, NGC925, NGC1232 and NGC1637.  The H{\sc ii} regions
have been splitted by density, which was derived from the [SII] ratio.
For seven regions, ion weighted temperatures from optical forbidden
auroral to nebular line ratios were obtained, and for six of them
oxygen abundances were derived using empirical calibration
methods. For the rest of the regions, metallicities have been
estimated from the S$_{23}$ calibration \citep{pm06} and tailored
photoionization models.

\cite{gv97} give data of 4 H{\sc ii} giant circumnuclear regions of
NGC~7714. As usual, densities were derived from the [SII]
lines. Oxygen abundances were obtained by the electron temperature
method thanks to the direct detection of the [OIII]$\lambda$4363 line.

\cite{zar94} made an analysis of 159 H{\sc ii} regions in 14 spiral
galaxies, from which we have chosen those that have well measured
[SII]]$\lambda\lambda$6717,31 and [SIII]]$\lambda\lambda$9069,9532
emission lines to obtain the density and the oxygen abundance through
the S$_{23}$ parameter \citep{pm06}. This reduces the number of HII
regions to 36.

We have also taken data from \cite{vzee06}, corresponding to 67 H{\sc
ii} regions in 21 dIrr. They provide the emission line intensities of
the [OII]$\lambda$3727, [OIII]$\lambda\lambda$5007,4959 and
[SII]$\lambda\lambda$6717,31 lines. In addition they also provide the
[SII] line ratio, which we have used to derive the electron
density. The [OIII] lines have been used to obtain the electronic
temperature and consequently the oxygen abundance.

Data from \cite{izo04} consist of H{\sc ii} regions in 76 blue compact
dwarfs (BCD) galaxies whose abundances have been derived using
electron temperatured from the ratio of the nebular to auroral lines
[OIII] lines. All the regions of this sample have oxygen abundances
lower than 12 + log(O/H) = 8.5. Data have been separated by density by
means of the [SII] line ratio.

Finally, \cite{yin07} provide data for 531 galaxies and H{\sc ii}
regions from the SDSS-DR4 sample, from which the [OIII]$\lambda$4363
line has been measured and the oxygen abundances have been derived by
standard methods.  These data do not include the [SII] lines and hence
no electron density could be derived.  The different observational
data sources and the method used to derive the oxygen abundances are
summarized in Table~\ref{referencias}.

\begin{table*}
\caption{Observational data used for this work}
\begin{tabular}{ccc}
\hline
Reference & Objects & [O/H] method \\
\hline
\citet*[][ and references therein]{diaz07} & High and low metallicity regions, CNSFR and
H{\sc ii} galaxies & Direct T$_{e}$ method and S$_{23}$/O$_{23}$\\
\cite*{cas02a} & high metallicity H{\sc ii}
regions& Direct T$_{e}$ method and S$_{23}$\\
\cite*{gv97}& H{\sc ii} circumnuclear regions&
Direct T$_{e}$ method \\
\cite*{zar94} & H{\sc ii} regions in Spirals& S$_{23}$\\
\cite*{vzee06} & H{\sc ii} regions in dIrr&Direct T$_{e}$ method \\
\cite*{izo04} & H{\sc ii} regions in BCD& Direct T$_{e}$ method \\
\cite*{yin07} &H{\sc ii} regions&Direct T$_{e}$ method \\
\hline
\label{referencias}
\end{tabular}
\end{table*}

\begin{figure*}
\resizebox{0.70\hsize}{!}{\includegraphics[clip,angle=0]{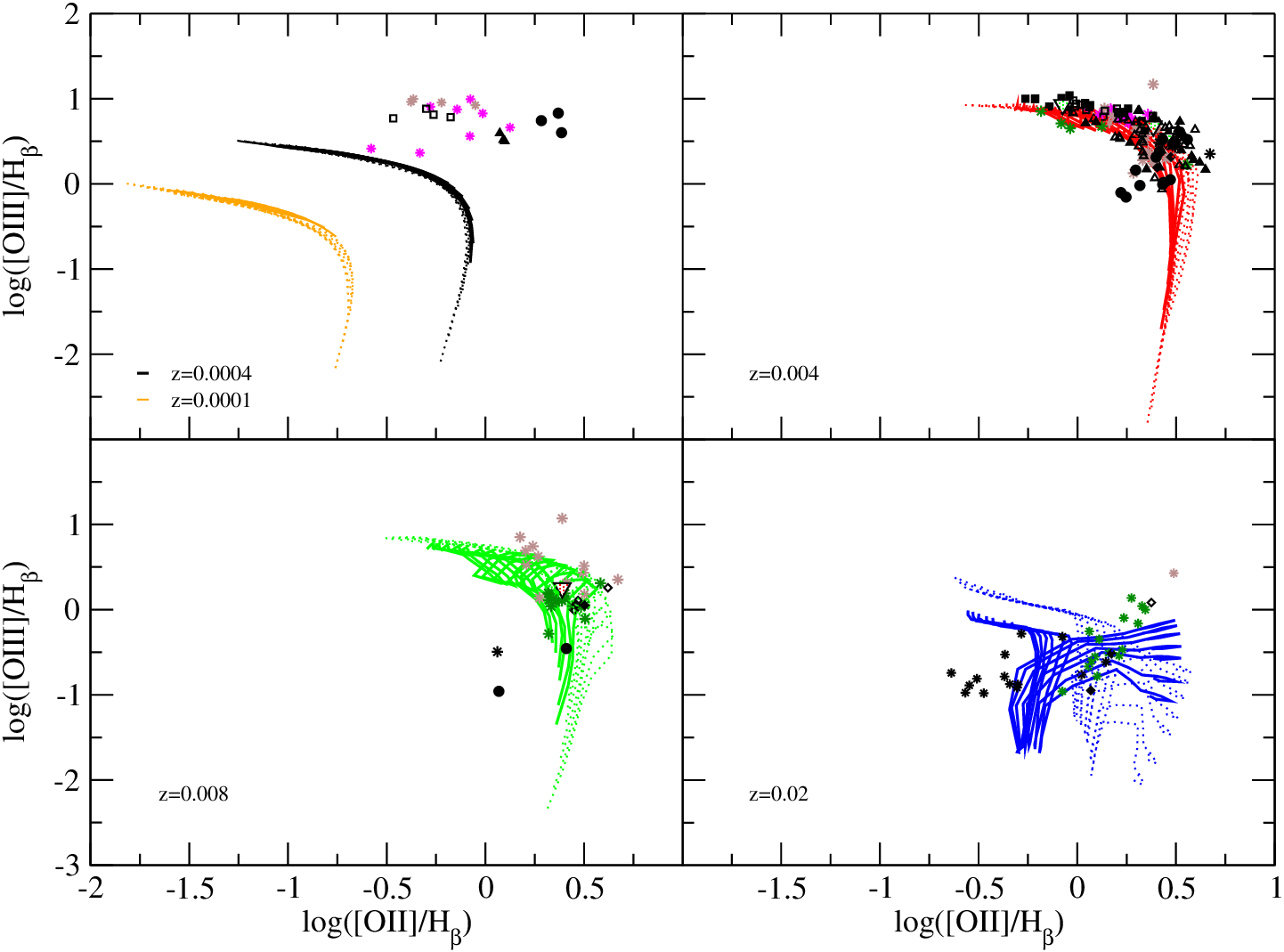}}
\caption{Diagnostic diagram showing
log([OIII]$\lambda\lambda$5007,4959, /H$_{\beta}$) {\sl vs}
log([OII]$\lambda$3727/H$_{\beta}$) for models computed with SAL2
IMF. Each panel shows models for a different metallicity and the
corresponding observational data in that metallicity
range. Observational data come from: \citet{zar94} (circles),
\citet{izo04} (squares), \citet{cas02b} (diamonds), \citet{vzee06}
(triangles up), \citet{gv97} (red triangles down), and \citet{diaz07}
(asteriscsof different colours: black: CNSFR; green: high metallicity
HII regions; brown: low metallicity H{\sc ii} regions; magenta: HII
galaxies). Filled symbols represent objects with $\rm n_{H} = 100
cm^{-3}$, while open symbols have $\rm n_{H} = 10 cm^{-3}$, for data
whose density has been obtained from the [SII] ratio.}
\label{lo2o3}
\end{figure*}

\subsection{Diagnostic Diagrams: optical emission line ratios}

We have plotted together the model results and the observational data
in different diagnostic diagrams in figures from \ref{lo2o3} to
\ref{ln2o3}. These diagrams can be used to study the relationship among
different emission line ratio and to extract information about the
physical parameters of the ionizing clusters.

Each figure is divided in 4 panels, one per metallicity: Z $=$ 0.0001
and 0.0004 (top left panel), 0.004 (top right panel), 0.008 (bottom
left panel) and 0.02 (bottom right panel). The highest metallicity
value of the grid, Z $=$ 0.05, has been excluded due to the absence of
spectroscopic data with those high abundances. The lowest metallicity
model tracks are plotted in the same panel as those corresponding to
Z=0.0004 to compare the position in the diagram of the emission lines
of such low metallicities. Solid and dotted lines represent
models with n$_{H}$=10 cm$^{-3}$ and n$_{H}$=100 cm$^{-3}$
respectively. In each panel the model metallicity is the central
value of the different ranges, except for ranges 1 and 4 that include
data with metallicities lower than Z $=$ 0.0004 and higher than Z $=$
0.02 respectively. {Different line tracks correspond to the evolution
of the models with different cluster masses, from the lowest one on
the right to the highest on the left. The observational data, taken
from different sources as explained in the previous section, whose
derived metallicities are in therange of the models we want to compare
with, are shown in each panel.  They have been divided by density and
subdivided by metallicity according to the ranges given in
Table~\ref{ranges}.

\begin{table}
\caption{Abundance ranges for the comparison of models and observational data.}
\begin{tabular}{cccc}
\hline
Label & Observed Metallicity Range  & Z in plotted models  \\
\hline
1     & 12+log(O/H) $<$  7.7        & 0.0004 \\
2     & 7.7 $<$ 12+log(O/H) $<$ 8.4 & 0.0040 \\
3     & 8.4 $<$ 12+log(O/H) $<$ 8.7 & 0.0080 \\
4     & 8.7 $<$ 12+log(O/H)         & 0.0200 \\
\hline
\label{ranges}
\end{tabular}
\end{table}

\begin{figure*}
\resizebox{0.70\hsize}{!}{\includegraphics[clip,angle=0]{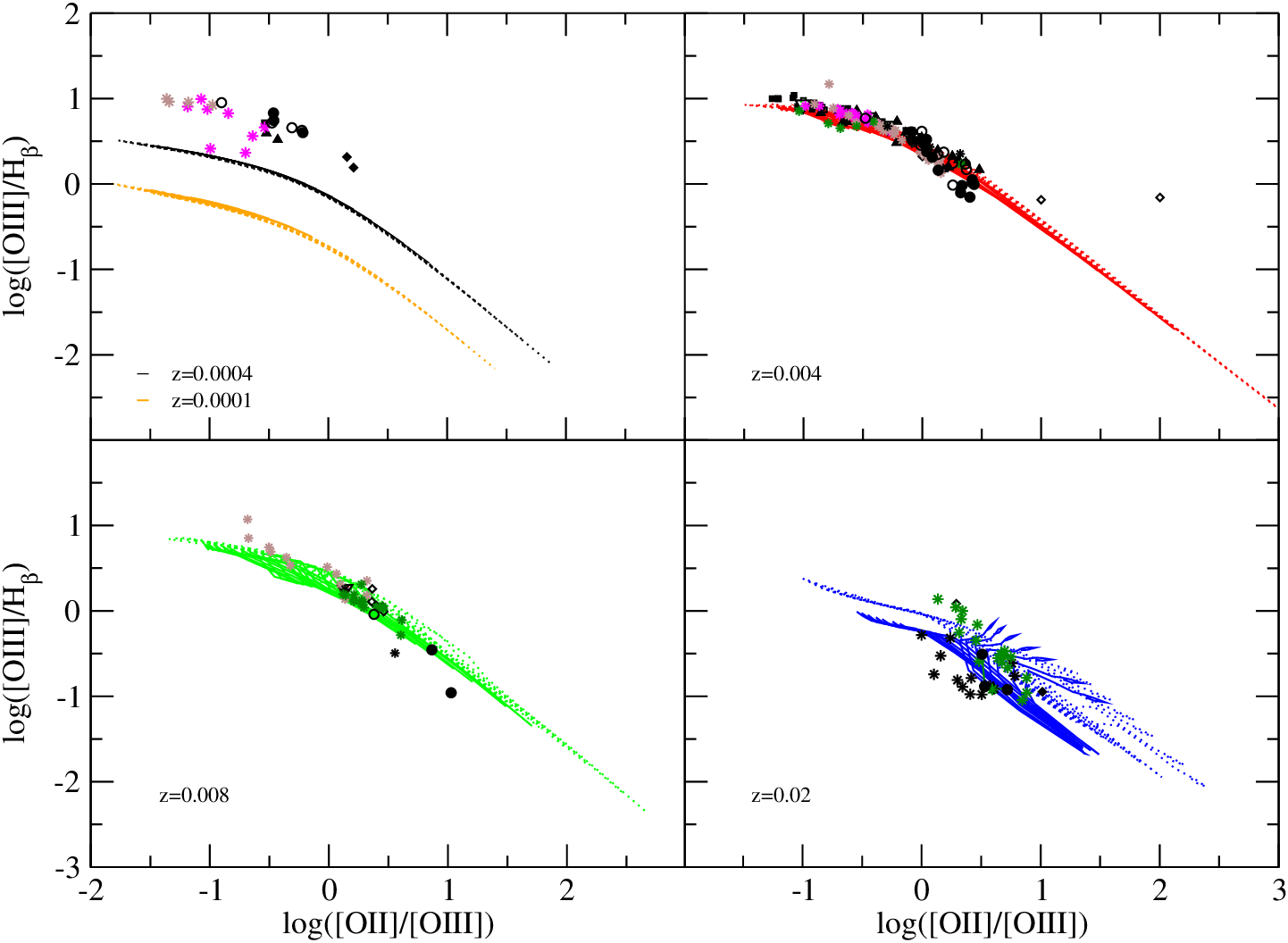}}
\caption{Diagnostic diagram showing
log([OIII]$\lambda$5007,4959/H$_{\beta}$) {\sl vs}
log([OII]$\lambda$3727/[OIII]$\lambda$,$\lambda$5007,4959) for models
computed with SAL2 IMF.  Each panel shows a different metallicity
model and the corresponding data in that metallicity range.  Symbols
have the same meaning as in Fig.~\ref{lo2o3}.}
\label{lo23o3}
\end{figure*}

\begin{figure*}
\resizebox{0.70\hsize}{!}{\includegraphics[clip,angle=0]{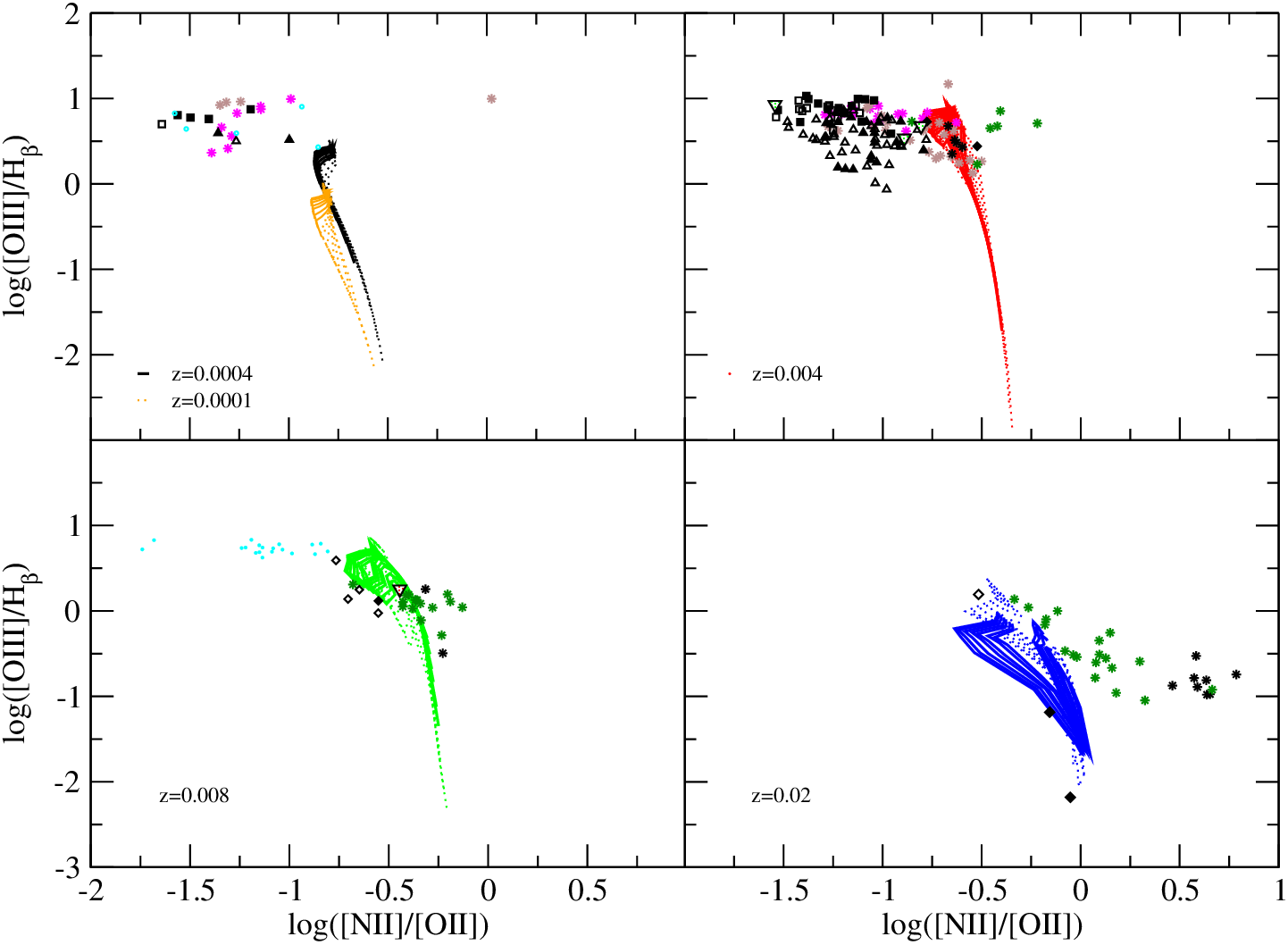}}
\caption{Diagnostic diagram showing
log([OIII]$\lambda\lambda$5007,4959/H$_{\beta}$) {\sl vs}
log([NII]$\lambda$6584/[OII]$\lambda$3727) for models computed with
SAL2 IMF. Each panel shows a different metallicity model and the
corresponding data in that metallicity range. Symbols have the same
meaning as in Fig.~\ref{lo2o3}, with the addition of the blue dots,
which correspond to H{\sc ii} regions from \citet{yin07}.}
\label{lno2o3}
\end{figure*}

\begin{figure*}
\vspace*{1.0cm}
\resizebox{0.70\hsize}{!}{\includegraphics[angle=0]{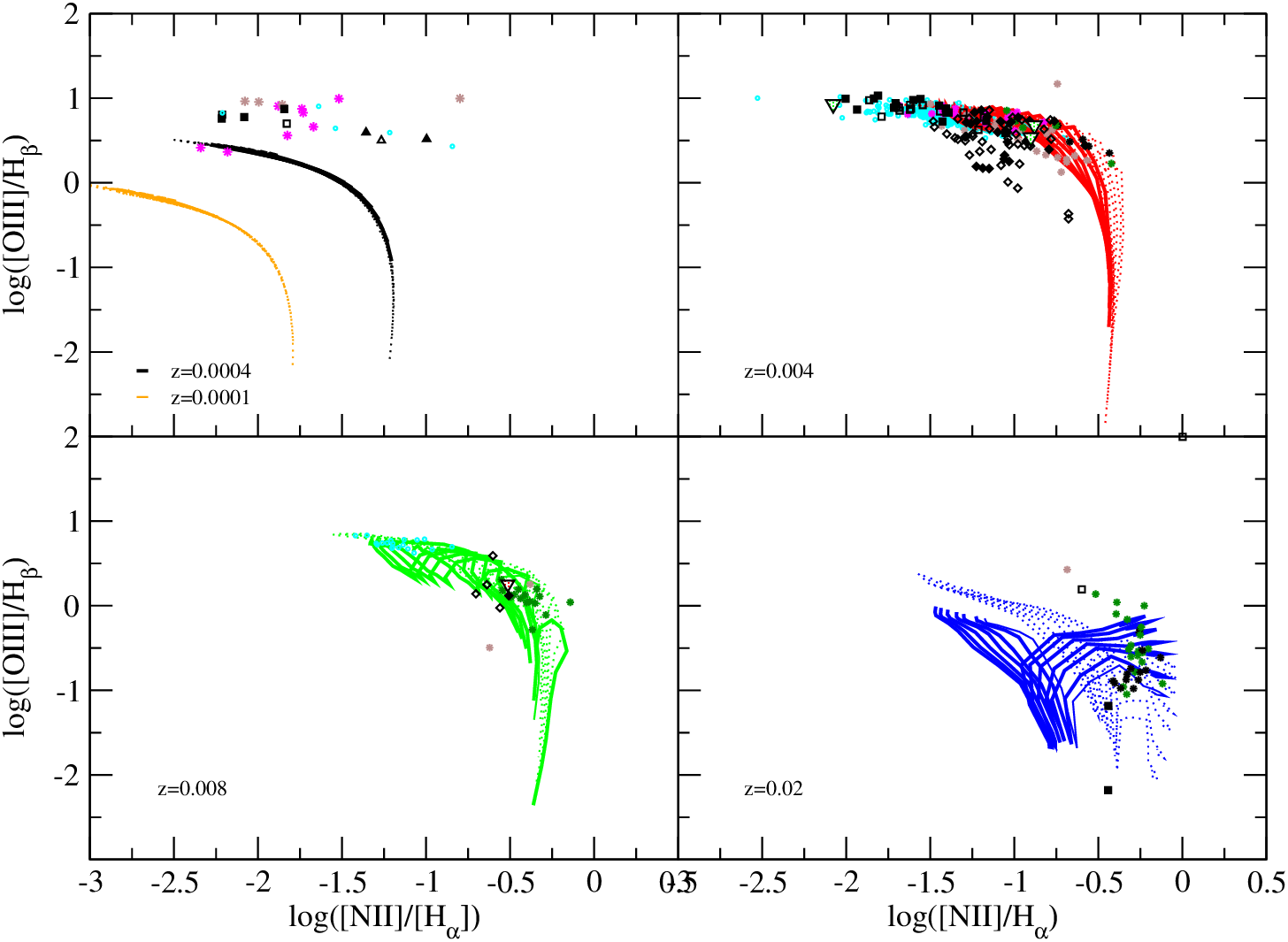}}
\caption{Diagnostic diagram showing
log([OIII]$\lambda\lambda$5007,4959/H$_{\beta}$) {\sl vs}
log([NII]$\lambda$6584/H$_{\alpha}$) for models computed with SAL2
IMF.  Each shell shows a different metallicity model and the
corresponding data in that metallicity range. Symbols have the same
meaning as the same than in Fig.~\ref{lo2o3} with the addition of blue
dots which correspond to H{\sc ii} regions from \citet{yin07}. }
\label{ln2o3}
\end{figure*}

The H{\sc ii} regions plotted here describe a well defined sequence in
diagnostic diagrams which involve oxygen line ratios, like
log([OIII]/H$_{\beta}$) vs log([OII]/H$_{\beta}$) and
log([OIII]/H$_{\beta}$) vs log([OII]/[OIII]) ( Fig.~\ref{lo2o3} and
Fig.~\ref{lo23o3}).  Most of the observed regions are located in the
panels corresponding to Z=0.004 and Z=0.008 since few data exist for
very low or oversolar metallicities. In fact, this is a bias effect
since most popular H{\sc ii} regions have been detected and catalogued
through their high intensities of the [OIII]$\lambda\lambda$5007,4959
lines, which happen in medium-low metallicity.  Furthermore, it is at
these metallicities, Z=0.004 and Z=0.008, that a direct measurement of
the [OIII]$\lambda$4363 line, used to derive the oxygen abundance, is
possible.
\begin{figure*}
\resizebox{0.70\hsize}{!}{\includegraphics[clip,angle=0]{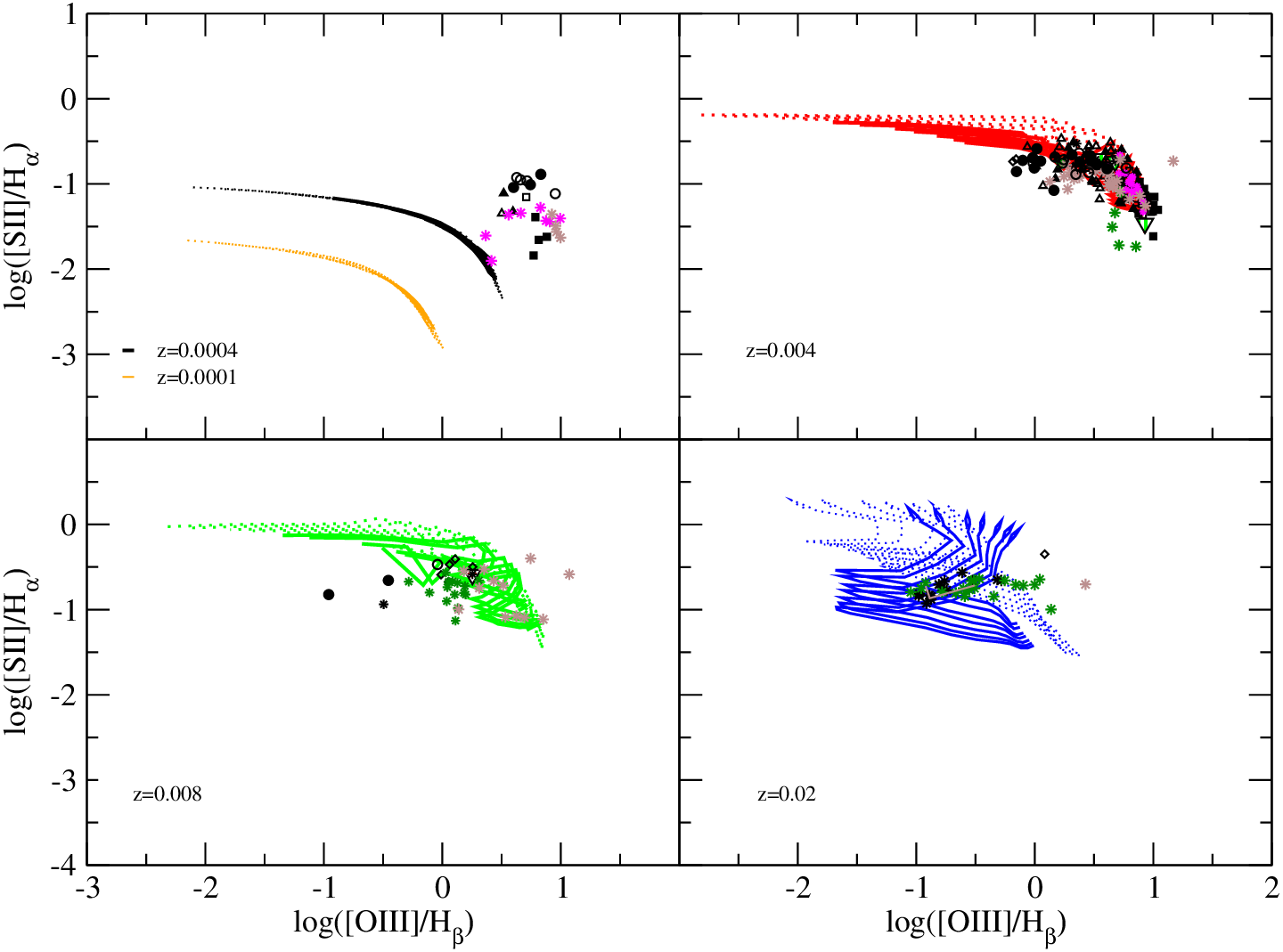}}
\caption{Diagnostic diagram showing
log([SII]$\lambda$6717,6731/H$_{\alpha}$) {\sl vs}
log([OIII]$\lambda\lambda$5007,4959/H$_{\beta}$) for models computed
with SAL2 IMF. Each shell shows a different metallicity model and the
corresponding data in that metallicity range . Symbols have the same
meaning as in Fig.~\ref{lo2o3}.}
\label{ls2o3}
\end{figure*}

Models of Z = 0.0001, 0.0004 and Z = 0.02 do not have the aim
of reproducing exactly the observation sequence but should be understood as
metallicity limits for the observed regions. Observational data in these
panels correspond to regions with metallicities lower than 12+log(O/H)
= 7.7, in the first panel (upper-left), or higher than 12+log(O/H) = 8.7
in the last panel (bottom-right), as summarized in
Table~\ref{ranges}.

 There are no observational data with Z $<$ 0.0004.  These regions
have very low values of the [OIII]$\lambda$4363 line, undetectable
with moderate size telescopes and standard S/N values, and therefore
their abundances have not been derived.

Our models predict that very low metallicity regions have small sizes
in H$\alpha$ images where most of these H{\sc ii} regions are
catalogued and therefore observational samples can be seriously biased
towards medium to large size H{\sc ii} regions with metallicities Z
$>$ 0.001 (see Section 4.4 and fig. \ref{evol_4363} for a detailed
discussion).  For high metallicities, the [OIII]$\lambda$4363 line is
not strong enough to be detected at a level of T$_{e}$ determination
and abundances have been derived through alternative methods as
already described.

Our models cover the observational data range corresponding to
different cluster masses. Different masses imply different numbers of
ionizing photons and hence different values of the ionization
parameter. These values decrease with time as the clusters evolve.
The models involving the most massive stellar clusters have the
highest ionization parameters, due to their higher number of ionizing
photons at a given age. Less massive clusters follow a similar time
evolution to the most massive ones, but starting at a lower initial
value.

As explained before, regions with intense
[OIII]$\lambda\lambda$5007,4959/H$_{\beta}$ line ratios correspond to
the highest ionization stages. For a given cluster mass, as the
cluster evolves, the high excitation emission lines tend to decrease
while [OII]/H$_{\beta}$ increases. This is due to the change of the
ionization parameter with cluster evolution. For a very young region,
whose ionization is dominated by O-B stars which produce a great
amount of ionizing photons , the size is still small, and the
resulting ionization parameter is high. As the cluster evolves, the
region size increases due to the stellar winds while the number of
ionizing photons remains almost constant (or even decreases) with
respect to early phases.  This effect makes the ionization parameter,
hence the emission line intensities and in particular the oxygen
lines, to decrease.
 
This evolution influences also the data sample since early ages result
in more compact regions (therefore high surface brightness) with high
ionization parameters. Thus, the emission line intensities of the
[OIII]$\lambda\lambda$5007,4959 lines are higher, increasing their
detectability in a given sample obtained with the same instrumental
configuration. In fact, most of the derived ages for those regions
would be younger than 5 Myr, when they emit enough ionizing photons to
produce a detectable H$_{\alpha}$ luminosity and from those, most of
them would be biased towrads younger ages (less than 3.5 Myr).

In all Figs.~\ref{lo2o3} to ~\ref{ls2o3}, there are not observations
found in the tail of our models with high [OII]$\lambda$3727 and low
[OIII]$\lambda\lambda$5007,4959 intensities. These values imply very
low ionization parameters, that indicates more diffuse regions with
low excitation. Since in most samples only the brightest H{\sc ii}
regions have been selected, due to their conspicuous H$\alpha$
emission or their high values of the [OIII]$\lambda\lambda$5007,4959
line intensities the lowest excitation regions tend to be
excluded. Some of these regions could also be older low metallicity
regions (in the age range from 5 to 10 Myr, see Figs.
~\ref{evol_lines_extendedz0001} and ~\ref{evol_lines_extendedz0004}).

The influence of the electron density, although of second order,
should also be considered. The models with n$_{H}$=100 cm$^{-3}$ show
lower values of the ionization parameter.  However, the radiation has
more particles per cm$^{-3}$ to ionize, and the excitation degree is
higher, producing model sequences shifted upwards from the lower
density ones.  More significant differences between models of
different densities are found in the highest metallicity diagrams
(Z=0.02).

\subsection{Sulphur and nitrogen emission line ratios.}

Diagnostic diagrams involving log([NII]$\lambda$6584/H$_{\alpha}$) and
log([SII]$\lambda\lambda$6717,31/H$_{\alpha}$) ratios have the
advantage of using lines very close in wavelength, and therefore
virtually reddening-free.

In large star forming regions, like starburts galaxies, the
relationship between the excitation parameter
[OIII]$\lambda\lambda$5007,4959/H$_{\beta}$ and the metallicity
indicator [NII]$\lambda$6584/H$_{\alpha}$ gives us information about
the youngest populations plus the star formation rate \citep{ken94}.
At the beginning of the evolution H$_{\alpha}$ emission is intense and
comes from the ionizing photons produced by the most massive stars,
most of them still in the main sequence. This number of ionizing
photons increases with the cluster mass and decreases with
metallicity. As the galaxy evolves, the H$_{\alpha}$ emission line
decreases, while the [NII]$\lambda$6584 line increases because of the
gas ejection by stars.

The [NII]$\lambda$6584/[OII]$\lambda$3727 ratio was introduced as a
good metallicity indicator by \cite{dop00}. Nitrogen is a secondary
element, or, at least, it has a strong secondary component, which
makes the ratio [NII]/[OII] to increase with metallicity. Besides, the
value of the mean temperature of H{\sc ii} regions decreases as
metallicity increases. When metallicity rises, the electron
temperature becomes too low to excite the O$^+$ transitions, hence
[OII] decreases while [NII] does not, thus increasing the [NII]/[OII]
ratio. A good correlation has been found between [NII]/[OII] and the
N$^{+}$/O$^{+}$ ionic abundance ratio, which is assumed to trace the
N/O ratio \citep{pm05}.  Also, \cite{diaz07} have shown that a very
tight correlation between [NII]/[OII] and N/O exists for high
metallicity H{\sc ii} regions and circumnuclear star forming regions,
which follow a sequence of increasing N/O ratio with increasing
emission line ratio.

In our diagnostic diagrams of Fig.~\ref{ln2o3} we can see how this
ratio changes with metallicity. For low metallicity models, both the
[OIII]/\hbeta\ ratio and [NII]/[OII] ratio are low. As metallicity
increases both line ratios also increase. When metallicity reaches the
limit of Z=0.004, [OIII] starts to decrease while [NII]/[OII] ratio is
still growing. At Z=0.02 the models show low values of [OIII] and high
values of [NII]. The only available observational data in this part of
the plot are high metallicity H{\sc ii} regions and CNSFR, showing
high values of [NII]/[OII], which indicate high N/O abundance ratios,
as can be seen in the sequence of \cite{diaz07}.

Regarding the nitrogen abundance, it is well known that N is mainly
primary for low metallicities and it is always possible to scale the N
abundance that we include as input in {\sc cloudy} by using a given
function, such as the one obtained by \cite{mvgd06} for the
relationship of N/O vs O/H. However, we remark that in the models
presented in this paper, we use SSP models where chemical evolution
has not been taken into account, either for N or for the other
elements as Fe or Ca. The nitrogen abundance has been scaled to the
solar value as it has been done for the rest of the elements. Each
element has its own evolution and in order to take this into account
it is necessary to make precise calculations following a consistent
method. This method can be seen in our previous works \citep{mmm08a,
mmm09phd} where we preent photoionization models that ise
as imput abundances the results obtained from chemical evolution
models.

Finally, Fig.~\ref{ls2o3} shows the
[SII]$\lambda\lambda$6717,31/H$_{\alpha}$ ratio {\sl vs}
[OIII]$\lambda\lambda$5007,4959/H$_{\beta}$ also almopst
reddening-independent.  We can see that high values of the
[SII]$\lambda\lambda$6717,31 line imply a low ionization parameter,
and therefore a low [OIII] emission line intensity. However, the
[OIII]$\lambda\lambda$5007,4959 lines are very metallicity dependent,
and a low intensity of this emission line could be due to a high
metallicity rather than to a low ionization parameter produced by an
evolved population. Although both ratios are indicators of the
ionization parameter, the values of \textit{u} derived from oxygen
emission line ratios can be lower than \textit{u} the values derived
from sulphur emission line ratios, which are not so metallicity
dependent.  

\subsection{Abundance determinations in H{\sc ii} regions and model consistency}

Emission lines allow the  determination of abundances by a direct method
based on the electronic temperature, T$_{e}$, derived from suitable emission lines from which those of [OIII] are the most popular. 
We have obtained oxygen abundances from the modelled emission lines in
order to check which is the most effective method to recover the oxygen
abundance used as input for the models. An accurate analysis requires
the detection of the auroral line [OIII]$\lambda$4363 with an intensity
larger than 5 per cent of H$_{\beta}$.

Fig.~\ref{evol_4363} shows the evolution of the [OIII]$\lambda$4363
/H$_{\beta}$ ratio for each metallicity, n$_{H}$=10, and for cluster
masses of 0.12, 0.20,0.4, 0.6 and 1$\times$ 10${^5}$ M$_{\odot}$.
This ratio is high for Z=0.0004 and Z=0.004, close to 5 per cent of
H$_{\beta}$ in the first Myr of the evolution. However, for both,
Z=0.0001 and Z=0.008, [OIII]$\lambda$4363 /H$_{\beta}$ has comparable
values and below standard detection limits.  Thus, the non detection of
this line can indicate either a high metallicity or a low oxygen
content.  For Z=0.02 there is no [OIII]$\lambda$4363 line produced for
this density case.

\begin{figure}
\resizebox{0.90\hsize}{!}{\includegraphics[clip,angle=0]{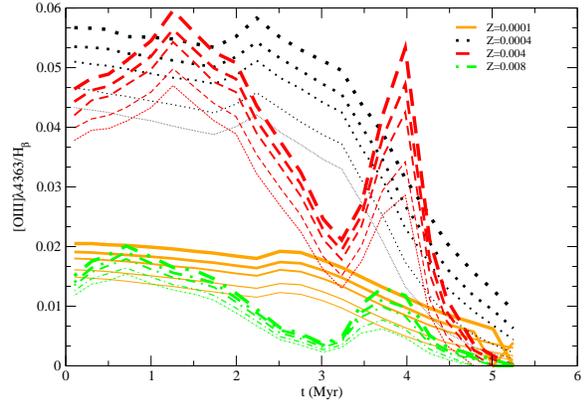}}
\caption{Evolution of the [OIII]$\lambda$4363 /H$_{\beta}$ ratio for
each metallicity, cluster masses from 0.12 to 1$\times$ 10${^5}$
M$_{\odot}$ and n$_{H}$=10.}
\label{evol_4363}
\end{figure}

Fig.~\ref{comp_oh} shows the comparison between model oxygen
abundances and the ones obtained by empirical methods. It can be seen
that, from our photoionization models, we obtain the auroral line
[OIII]$\lambda$4363 along the whole cluster evolution for
metallicities in the range 0.0001$<$ Z $<$ 0.008. The [OIII] line
electron temperature method gives good results for all of them, with
differences lower than 0.05 dex between input and derived abundances.

\begin{figure}
\resizebox{0.90\hsize}{!}{\includegraphics[clip,angle=0]{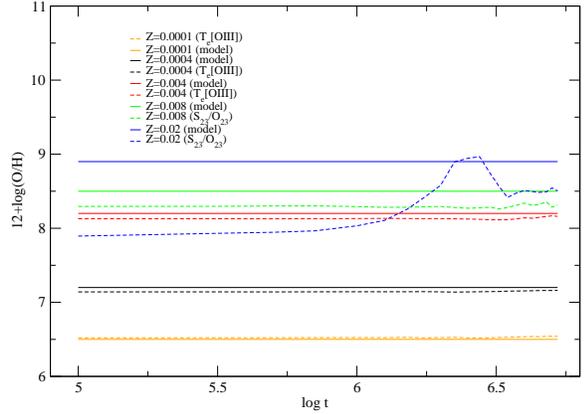}}
\caption{Comparison between abundances used as input for the models
(solid lines) of different metallicities and abundances obtained
using different empirical calibrators (dashed lines).}
\label{comp_oh}
\end{figure}

At Z $=$ 0.008, corresponding to an oxygen abundance of 12+log(O/H)
=8.5, the [OIII]$\lambda$4363 line intensity range goes from 1.62 per
cent of the H$_{\beta}$ intensity 1 Myr after the star formation, to
0.82 per cent 4 Myr after the burst.  However, despite this line being
detectable at this metallicity, the electronic temperature method is
not able to reproduce the input metallicity of the model.  Simple
photoionization models are not able to reproduce the derived T$_{e}$
in some situations.  \cite{stas78,stas02,stas05} showed that for high
metallicity H{\sc ii} region models, the oxygen abundance derived from
predicted emission line ratios and the T$_{e}$-method is lower than
the one used as input in the models}. These differences may be
significant when large temperature gradients or fluctuations are
present. In these cases, the temperature obtained from the emission
line, T$_{l}$[OIII], is different from that derived from the [OIII]
line ratio, T$_{r}$, and from the average ionic temperature
T(O$^{++}$), with T$_{l}$[OIII] and T$_{r}$ obtained from the models
not being the representative temperature of the high ionization
O$^{++}$ zone.  Fig.~\ref{oh_oh} shows a plot similar to Fig.1 of
\cite{stas05} comparing a sequence of model derived oxygen abundances
as a function of the input values . In this diagram, we can see that
both sets of values coincide for low metallicities; for metallicities
around 8.6 however, deviations become notorious and for higher
metallicities the deviations are much more significant.

The abundances analysis in high metallicity H{\sc ii} regions is
hampered by the fact that oxygen optical lines act as main coolant for
the nebula, hence a higher oxygen abundance leads to a more effective
cooling. As the gas cools down, the electron temperature gets lower
and the optical [OIII] forbidden lines get weaker. In general, this
low excitation makes any temperature-sensitive lines in the optical
range too weak to be measured. For this reason empirical calibrators
involving other lines besides the optical oxygen ones should be used
\citep{pm05, diaz07}.  In our models, for metallicities Z $=$ 0.008
and Z $=$ 0.02, abundances have been determined with the ratio
S$_{23}$/O$_{23}$, obtaining reasonably good results for Z=0.008 (with
a difference of 0.2 dex approximately), but not for Z=0.02, which
shows important differences at the beginning of the evolution. For Z
$=$ 0.05 we cannot recover the oxygen abundance used as input by any
empirical method since an effective calibration for such high
metallicity range \citep[which is approximately 3Z$_{\odot}$ in the
reference system of][]{gre98} has not been yet devised.

\begin{figure}
\vspace*{1.0cm}
\resizebox{0.90\hsize}{!}{\includegraphics[angle=0]{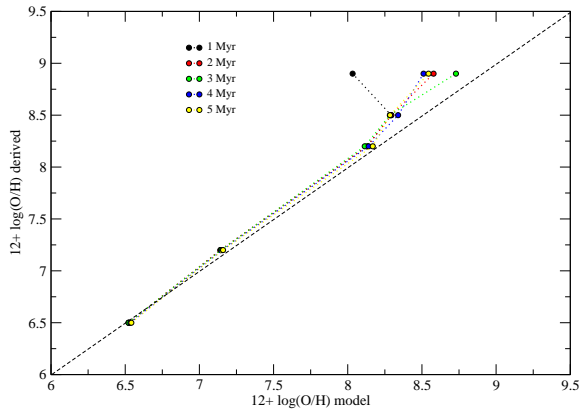}}
\caption{Comparison between abundances used as input for the models
and abundances obtained using T$_{e}$-based methods. Each color
corresponds to a different age sequence. The values of 12+log(O/H)
derived empirically are very close to the input values up to 8.6.}
\label{oh_oh}
\end{figure}

\subsection{Ionized Region equilibrium time and ionizing cluster age} 

The output of the photoionization models shows the picture of the
ionized region once the equilibrium state is reached. This means that,
when we observe a H{\sc ii} region, what we really see is the effect
of an ionizing cluster of a given age, regardless the actual age of
the ionized nebula. The time step in our models corresponds to the
stellar cluster age step given by the isochrone time resolution. The
consequence is that the age resolution with which we can date the
ionizing cluster from the information of the gas ionizing spectrum
depends on the speed at which the nebula reaches the equilibrium, and
this process is controlled by the metallicity. In the worst case, if
the equilibrium time is longer than the cluster's age, it would be
possible to see the effect of the ionizing cluster several Myrs after
the ionization. Therefore, the ionizing stellar cluster would be older
than inferred from the emission line spectrum, being this observed
cluster the result of its evolution during equilibrium time of the
nebula.  \cite{wof09} shows that the time it takes for a H{\sc ii}
region to reach thermal equilibrium at a metallicity of Z $=$ 0.001
can be longer than 1 Myr within the first 5 Myr of evolution. The
equilibrium time scale increases with decreasing hydrogen density. In
addition, it increases at lower metallicities, due to the low content
of metals for cooling the gas.

\begin{figure}
\vspace*{1.0cm}
\resizebox{0.90\hsize}{!}{\includegraphics[clip,angle=0]{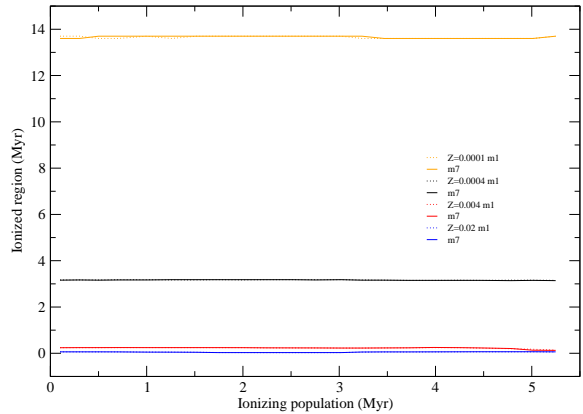}}
\caption{Relationship between the age of the ionizing stellar cluster
and the recombination age of the ionized region. Orange, black, red
and blue lines join models for Z $=$ 0.0001, 0.0004, 0.004 and 0.02
respectively. Solid lines correspond to models for a cluster mass of
to 1$\times$ 10${^5}$ M$_{\odot}$ while dotted lines correspon to a
cluster mass of 0.12$\times$10${^5}$ M$_{\odot}$ }
\label{nebula_star}
\end{figure}

\begin{figure*}
\resizebox{0.90\hsize}{!}{\includegraphics[clip,angle=90]{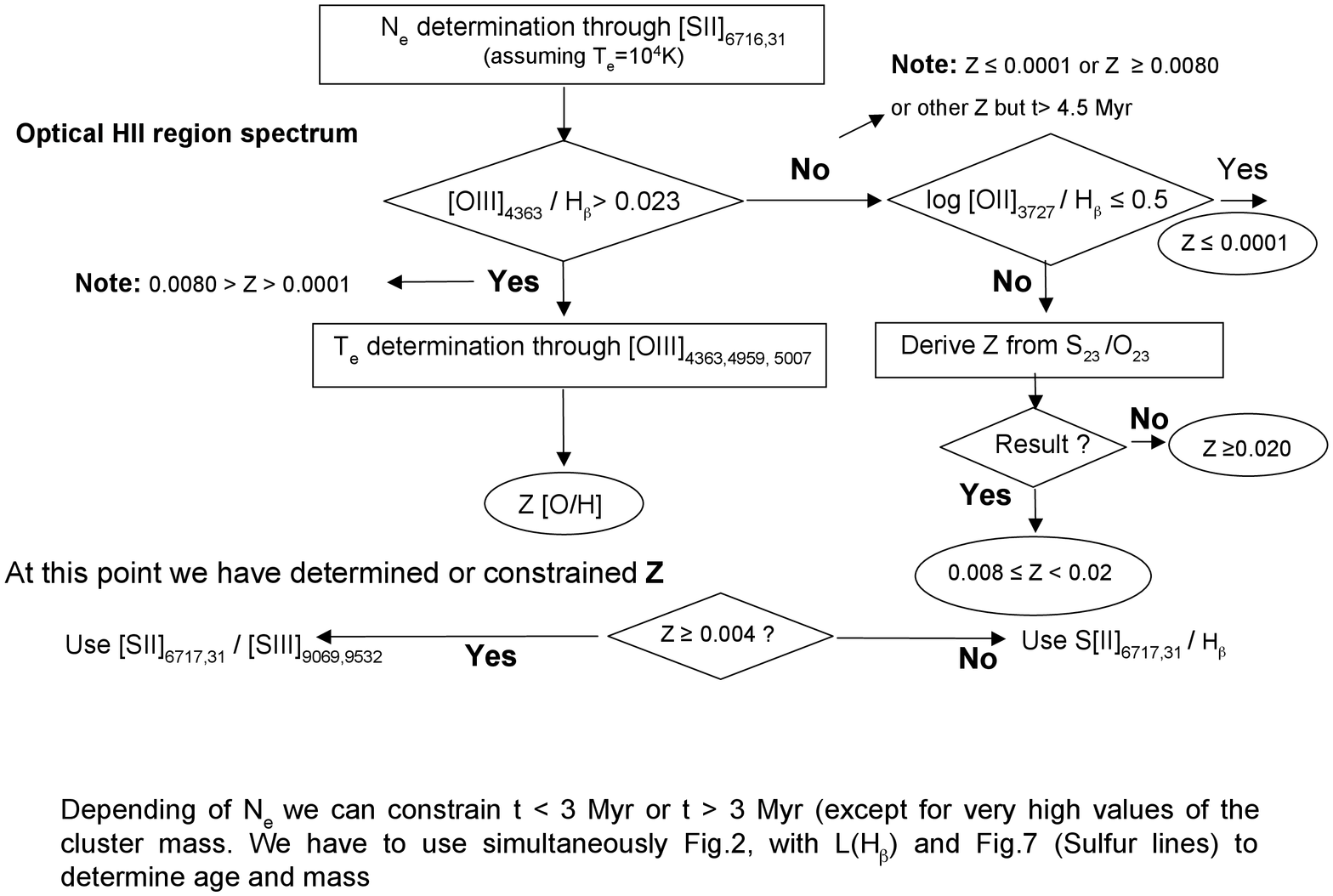}}
\caption{Flow Diagram to unveil the ionizing cluster physical
properties through the gas optical emission line spectrum}
\label{dia_flow}
\end{figure*}

We have tested this effect in our models by checking the equilibrium
time in each {\sc cloudy} model. Fig.~\ref{nebula_star} plots the age
of the ionizing region given by {\sc cloudy} once the equilibrium has
been reached versus the age of the ionizing star cluster.  Lines in
orange, black, red and blue correspond to models with Z $=$ 0.0001,
0.0004, 0.004 and 0.02 respectively. Solid lines refer to models for a
cluster mass of 1$\times$ 10$^5$\Msun\ while dotted lines to models
for a cluster mass of 0.12$\times$ 10$^5$ \Msun. All models have been
computed with a density of n$_{H}$=10~cm$^{-3}$.

A first conclusion to be extracted from this graph is that the age of
the ionized region does neither depend on the cluster mass nor on the
cluster age but it is controlled by the metallicity. This implies that
neither the total number of ionizing photons nor the ionizing spectrum
shape have a significant role in setting the equilibrium time of the
ionized regions, which is controlled by the electron temperature
(given by the metallicity). This is clearly seen in
Fig.~\ref{nebula_star} where dotted and solid lines overlap and the
ionized region age is almost constant along the cluster
evollution. According to the computed models, the average equilibrium
age is reached around t$_{eq}$= 13.6, 3.1, 0.2, 0.1 and 0.05 Myr for
metallicities Z $=$ 0.0001, 0.0004, 0.004, 0.008 and 0.02,
respectively.

The second conclusion is that there are two metallicity regimes in
which the ionizing cluster age has to be interpreted in a different
way from the nebular emission line spectrum. In moderate to high
metallicity H{\sc ii} regions (Z $\geq$ 0.004) the ionized nebula
equilibrium time is shorter than the ionizing cluster age by a factor
between 4 and 10. This means that for regions with metallicities above
0.004 we will derive the same age for the ionizing cluster whether we
look directly to the embedded cluster (whenever possible) or we
observe the cluster effects on the surrounding nebula, that is the
nebular emission line spectrum. In these cases, the resolution time,
and therefore the error with which we can determine this age, is
around 0.05 Myr at Z=0.02, 0.1 Myr at Z=0.008 and 0.2 Myr at
Z=0.004. In this metallicity regime, if we detect WR bumps in the
optical spectrum or O, B and WR stars from \textit{UV} images, we can
assure, according to our results, that these stars are the ones
responsible for the detected nebula ionization. On the contrary, in
low metallicity H{\sc ii} regions, there is a delay between the time
at which we observe the ionized region and the age of the stellar
cluster responsible for this ionization. At Z = 0.0004 this delay is
of the order of 3 Myr, what is consistent with the value of 1 Myr
derived by \cite{wof09} for Z=0.001. The delay can be as long as 13
Myr at Z = 0.0001.  This implies that we may observe low metallicity
regions apparently ionized by a young cluster but without finding the
expected young and hot stars when we observe directly inside the
nebula. Moreover, we could find more evolved stars radiating in the
\textit{IR} as RSGs as the resulting products of the evolution of the
ionizing stars. We suggest to take into account this result when
interpreting the stellar content inside low metallicity H{\sc ii}
regions.

\subsection{Unveiling the physical properties of the ionizing clusters}

The purpose of this piece of work is to show that it is possible to
unveil the properties of the young cluster responsible for the
ionization on the basis of the emission line spectrum of the ionized
gas alone. In what follows we suggest some recommendations to find the
physical properties of the dominant young ionizing population (see
Flow Diagram in Fig.~\ref{dia_flow}).

\begin{itemize}

\item[--] The first parameter than can be determined is the
\textbf{electron density}, which is derived from the
[SII]$\lambda\lambda$6716, 6731 lines. This value influences the
ionization parameter and therefore it is an important input for our
models.

\item[--] The \textbf{metallicity} can be derived or constrained once
the electron density has been determined. If the value of
[OIII]$\lambda$4363 /H$_{\beta}$ is higher than 0.023, then abundances
can be derived by standard methods using the electron temperature
calculated from the [OIII]$\lambda$4363, [OIII]$\lambda$4959,
[OIII]$\lambda$5007 line ratio. We remark that the absence (or a very
low value) of optical [OIII] lines is an indicator of very low or very
high metallicity. PopStar includes models with metallicity as low as
Z=0.0001. If the value of [OIII]$\lambda$4363 /H$_{\beta}$ is lower
than 0.023, we recommend to check the value of log([OII]$\lambda$3727
/H$_{\beta}$). Values lower than 0.5 indicate very low metallicities
while the opposite indicates that the region is in the high
metallicity regime. In this latter case, we can estimate the
metallicity through empirical methods \citep{pm05}. These methods
allow to estimate abundances up to solar values. (o (Z = 0.02). Values
higher than those are usually not found in optically selected samples
since these regions are efficiently cooled by infrared lines and have
a very faint optical emission line spectrum.

\item[--] The \textbf{age} should be constrained once the metallicity
evaluation has been carried out. We propose to use the diagnostics
with [SII]$\lambda\lambda$6716,31/H$_{\alpha}$ (Fig.~\ref{s2t}), for
low metallicity regions, and
[SII]$\lambda\lambda$6716,31/[SIII]$\lambda\lambda$6069,9532
(Fig.~\ref{s23t}), for high metallicity regions. In both cases, we
need to use the electron density to select among different sets of
models. We have computed models for two density values: 10cm$^{-3}$
and 100 cm$^{-3}$, which we recommend to use when electron density is
less than 50cm$^{-3}$ or higher than 50cm$^{-3}$ respectively, and up
to a maximum value of 400 cm$^{-3}$. On the basis of these diagrams
and the observed values of the line ratios, we can distinguish if the
ionizing cluster is younger or older than 3 Myr. The resolution at
which we can estimate the age depends on the equilibrium time as
discussed in section 4.5. In the case of low metallicity regions
(below Z = 0.004) the equilibrium time is of the order, or even longer
than the ionizing cluster age. This has to be taken into account when
observations of the embedded stars are available, since in this low
metallicity scenario these stars are the result of the evolution of
the ionizing cluster and not the ionizing cluster itself.

\item[--] The luminosity of \halpha\ emission line is directly related
to the number of ionizing photons, and therefore to the mass of the
cluster. However, \halpha\ luminosity depends not only on the cluster
mass but also on the age and metallicity, even during the first 5 Myr
of evolution (see Fig.~\ref{evol_lines}) when the emission line
spectrum in the optical is detectable, according to our
models. Moreover \halpha\ is detected up to an age of 20 Myr, when the
emission line spectrum is not yet detectable (we remind the reader
that this paper is focused on the first few Myr when emission lines
are prominent since these are the regions available in most H{\sc ii}
region catalogues). Due to the former reasons, it is difficult to
derive the mass without previous estimates of metallicity and
age. Therefore we propose to derive the \textbf{mass} of the cluster
from the H$_{\alpha}$ luminosity (Fig.~\ref{evol_lines}) using the
previously found restrictions on metallicity and age.

\item[--] Finally, we can use the values of the complete set of emission lines to check the consistency of our determination. 

\end{itemize}

Since this process is slow to be applied to a large sample of objects,
we are developing a fast software tool, based on the previous
guidelines and using minimization algorithms, to find the best model
that fits the emission line values and determine the physical
properties of the young cluster responsible for the ionization of a
given H{\sc ii} region, when spectroscopic data are available.

\section{Summary and conclusions}

We present a new grid of evolutionary models as a tool to estimate the
mass, age and metallicity of the dominant ionizing stellar population
of a given H{\sc ii} region from its optical emission line spectrum.

We provide a complete set of photoionization models for H{\sc ii}
regions along the first 5.2 Myr of their evolution, using as ionizing
source a very young star cluster obtained from the theoretical
evolutionary synthesis models PopStar \citep{mgb09}. \footnote{ More
evolved cluster emission line spectra can be provided under request,
however, these spectra will have only low ionization emission lines
and in particular do not present [OIII] lines.}  The basic grid is
composed by SSPs computed with different IMFs and a wide range
of metallicities (Z=0.0001, 0.0004, 0.004, 0.008, 0.02 and 0.05). A
Salpeter IMF with m$_{low}$=0.15M$_{\odot}$ and m$_{up}$ =
100M$_{\odot}$ has been used in this work. The ionizing cluster is
assumed to form in a single burst. Our grid considers five values of
the total cluster mass (0.12, 0.20, 0.40, 0.60, 1.00, 1.50 and
2x10$^{5}$M$_{\odot}$), covering the range of observed H{\sc ii}
regions.  As the cluster evolves, the stellar winds and supernovae
blow the material, determining the H{\sc ii} region size. This
material is ionized by the emergent continua. We use the
photoionization code {\sc cloudy} to obtain the corresponding H{\sc
ii} region emission line spectrum once the equilibrium is reached.

We present diagnostic diagrams where we compare our model results with
a large sample of observed H{\sc ii} regions, whose metallicities have
been consistently computed in a homogeneous manner according to the
most appropriate method for each metallicity range. It is found that,
in general, our models can satisfactorily reproduce the observations.

Finally, we show that the emission line spectrum from an ionized
region provides a powerful tool to unveil the physical properties of
the embedded young cluster responsible for the gas ionization and
propose some guidelines to constrain its physical properties.

\section{Acknowledgments}

This work has been partially supported by DGICYT grant
AYA2007--67965-C03-03 and partially funded by the Spanish MEC under
the Consolider-Ingenio 2010 Program grant CSD2006-00070: First Science
with the GTC (http://www.iac.es/consolider-ingenio-gtc/).  Also,
partial support from the Comunidad de Madrid under grant
S-0505/ESP/000237 (ASTROCAM) is acknowledged. The authors thank an
anonymous referee for valuable comments and suggestions that have
contributed significantly to the quality of this paper.

\appendix
\label{appA}
\section{Ionization parameter recalibration.}

This appendix includes a re-calibration of the ionization parameter
for low and high metallicity H{\sc ii} regions with sulphur lines. As
we have explained in the paper, we do not use the ionization parameter
as an input for our models since we derive it from the physical
properties of the ionizing clusters and the assumed geometry of the
ionized nebula. However, since many other models use this magnitude to
parametrize their grids or use different geometries for the ionized
regions, we have considered interesting to obtain a re-calibration of
the relationship between the ionization parameter and the sulphur lines. We
have used for that the results from our models, as a
self-consistent set of values typical for H{\sc ii} regions.


The [SII]$\lambda\lambda$6717,31/[SIII]$\lambda\lambda$9069,9532 line
ratio has been shown to be a good ionization parameter indicator for
moderate to high metallicities \citep{diaz91}, with little dependence
on metallicity.  \cite{gbd95a}
proposed to use the ratio [SII]/[SIII] to derive the ionization
parameter of high metallicity regions (Z $>$ 0.004).  We have
re-calibrated this relationship for each model. This calibration has
been made for values of the ionization parameter logu~$<$~-2 since
most H{\sc ii} regions observational data do not show ionization
parameters higher than this. For a more accurate calibration
for higher ionization parameter values
a particular model, according to cluster mass, should be selected a high dispersion in the model
results for different masses is found. 

For high metallicity models we
find a cubic polynomial relationship between log u and log[SII]/[SIII].
There are small differences between models of Z=0.008 and Z=0.004 and 
models of Z=0.02, which makes necessary to provide two different
equations for log~u in order to obtain accurate calibrations.

For \textbf{n$_{H}$=10}, we obtain:

\textit{logu = -2.865 -1.175x +0.139x$^{2}$ -0.135x$^{3}$} (Z=0.008,  Z=0.004)

\textit{logu = -2.705 -1.736x +0.698x$^{2}$ -0.973x$^{3}$} (Z=0.02)

where \textit{x = log([SII]$\lambda\lambda$6717,31/[SIII]$\lambda\lambda$9069,9532)}


For \textbf{n$_{H}$=100}, we obtain:

\textit{logu = -3.136- 0.810x +0.685x$^{2}$ -0.784x$^{3}$} (Z=0.008, Z=0.004)

\textit{logu = -2.938 -1.051x +0.043x$^{2}$ -0.482x$^{3}$} (Z=0.02)

where \textit{x = log([SII]$\lambda\lambda$6717,31/[SIII]$\lambda\lambda$9069,9532)}

\begin{figure}
\resizebox{0.90\hsize}{!}{\includegraphics[clip,angle=0]{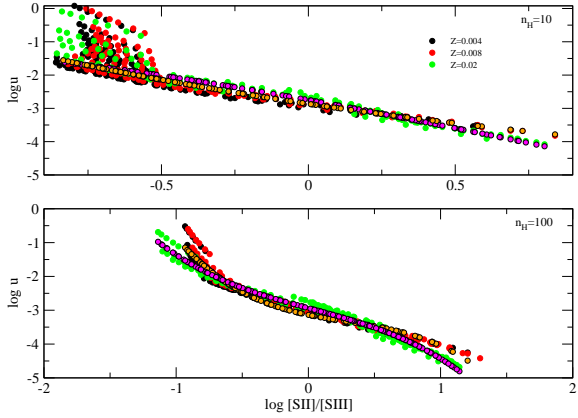}}
\caption{Calibration of the ionization parameter u from the [SII]$\lambda\lambda$6717,31/[SIII]$\lambda\lambda$9069,9532 line ratio for high metallicity models: Z=0.004 (black dots), Z=0.008 (red dots), Z=0.02 (green dots), and
two particle densities, n$_{H}$=10 cm$^{-3}$ (upper panel) and n$_{H}$=100
cm$^{-3}$(lower panel).  The calibration has been made for values of
log u$<$ -2 and it is represented by orange dots.}
\label{zhigh}
\end{figure}

For low metallicity models (Z $<$ 0.004) the calibration has been made
using log[SII]/H$_{\alpha}$, finding a cubic polynomial
relationship with log~u for values of
log~u$<$~-2.

For \textbf{n$_{H}$=10} the fittings are:

\textit{logu = 18.89 +33.00x +15.69x$^{2}$ +2.33 x$^{3}$} (Z=0.0001)

\textit{logu = -2.51 + 3.96 x +3.54 x$^{2}$ +0.69 x$^{3}$} (Z=0.0004)

where \textit{x=[SII]$\lambda\lambda$6717,31/H$_{\alpha}$}

For \textbf{n$_{H}$=100} the fittings give the following result:

\textit{logu = -19.95 -19.29x -7.56 x$^{2}$ -1.07x$^{3}$} (Z=0.0001)

\textit{logu = -11.75 -12.98 x -6.60 x$^{2}$ -1.26 x$^{3}$} (Z=0.0004)

where \textit{x=[SII]$\lambda\lambda$6717,31/H$_{\alpha}$}

\begin{figure}
\resizebox{0.90\hsize}{!}{\includegraphics[clip,angle=0]{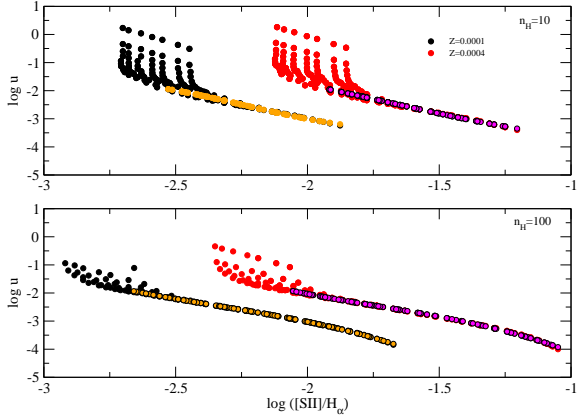}}
\caption{Calibration of the ionization parameter \textit{u} from the
[SII]/H$_{\alpha}$ line ratio for the models with metallicities
Z=0.0001 (black dots) and Z=0.0004 (red dots), and both densities:
n$_{H}$=10 cm$^{-3}$ (upper panel) and n$_{H}$=100 cm$^{-3}$ (lower
panel).  The calibration is represented by orange dots.}
\label{zlow}
\end{figure}

\end{document}